\documentclass{article}[12pt,psfig]
\usepackage{amsmath}
\usepackage{latexsym}
\usepackage{float}
\usepackage{amssymb}
\usepackage{graphicx}
\usepackage{epsfig}
\usepackage{cite}
\begin{document}

\title{Molecular Dynamics Simulations}

\author{%
 Kurt Binder$^{\text{1)}},\;$
 J\"urgen Horbach$^{\text{1)}},\;$\,%
 Walter Kob$^{\text{2)}}$,\\
 Wolfgang Paul$^{\text{1)}}$, and
 Fathollah Varnik$^{\text{3)}}$ \\[\baselineskip]%
                   $^{\text{1)}}$\textit{ Institut f\"{u}r Physik, Johannes Gutenberg-Universit\"{a}t,} \\
                   \textit {D-55099 Mainz, Staudinger Weg 7, Germany}\\%
                   $^{\text{2)}}$\textit{ Laboratoire des Verres, Universit\'e
                                   Montpellier II,}\\
                                  \textit{F-34095 Montpellier,
                                  France}\\
                   $^{\text{3)}}$\textit{ Centre de Calcul Atomique et Moleculaire (CECAM),  } \\
                   \textit { ENS-Lyon, 46, All\'ee d'Italie, F-69007 Lyon, France}\\%
\date{}
}
 \maketitle

\renewcommand{\baselinestretch}{1.3}

\begin{abstract}
A tutorial introduction to the technique of Molecular Dynamics
(MD) is given, and some characteristic examples of applications
are described. The purpose and scope of these simulations and the
relation to other simulation methods is discussed, and the basic
MD algorithms are described. The sampling of intensive variables
(temperature $T$, pressure $p$) in runs carried out in the
microcanonical ($NVE$) ensemble ($N$= particle number, $V$ =
volume, $E$ = energy) is discussed, as well as the realization of
other ensembles (e.g. the $NVT$ ensemble). For a typical
application example, molten SiO$_2$, the estimation of various
transport coefficients (self-diffusion constants, viscosity,
thermal conductivity) is discussed. As an example of
Non-Equilibrium Molecular Dynamics (NEMD), a study of a
glass-forming polymer melt under shear is mentioned.
\end{abstract}

\clearpage

\section{Introduction: Scope and Purpose of Molecular Dynamics
Simulations}
\subsection{Molecular Dynamics and its Relation to Other Methods
of Computer Simulation}

Computer simulations in condensed matter physics aim to calculate
structure and dynamics from atomistic input \cite{1,2,3,4}. The
theoretical basis of this approach is statistical thermodynamics.
The conceptually simplest approach is the classical Molecular
Dynamics (MD) method \cite{5,6,7}: one simply solves numerically
Newton's equations of motion for the interacting many particle
system (atoms or molecules interacting, e.g., with pair potentials).
The basics of the method thus is nothing but classical mechanics,
and one creates a deterministic trajectory in the phase space
of the system. Now the idea is to simply take time averages of the
observables of interest along this trajectory, and rely on the
ergodicity hypothesis of statistical mechanics, which asserts that
these time averages are equivalent to ensemble averages of the
appropriate microcanonical ($NVE$) ensemble. Of course, Newton's
equations of motion conserve the total energy $E$, and hence the
conjugate intensive thermodynamic variables such as temperature
$T$ and pressure $p$ can only be inferred indirectly and exhibit
fluctuations (since the particle number $N$ is finite and
sometimes even fairly small, such fluctuations must not be
neglected and need careful consideration). Sometimes it is
advantageous to directly realize other ensembles of statistical
mechanics, such as the constant volume $V$- constant temperature
$T$ ($NVT$) ensemble or the $NpT$ ensemble, and --- as we shall see
later --- this is indeed feasible by introducing a coupling to
appropriate ``thermostats'' or ``barostats''.

An alternative way of carrying out a MD simulation at constant temperature 
is possible by introducing an artificial weak friction force,
together with random forces whose strength are controlled by the
fluctuation-dissipation theorem. Such techniques are very common
e.g.~for the simulation of polymer melts \cite{8,9}. This method
is very closely related in spirit to stochastic simulation methods
such as ``Brownian Dynamics'' where one simulates a Langevin
equation (the inertial term in the equation of motion being
omitted). While dynamical correlations for such methods differ
somewhat from strictly microcanonical MD methods, for the
computation of static properties from averages along the
stochastic trajectory in phase space such methods can be
advantageous.

This statement is also true for the importance sampling Monte Carlo
(MC) method \cite{10,11}. As is well known \cite{4,11}, MC sampling
means that one creates a random walk-like trajectory in configuration
space, controlled by transition probabilities that ensure the approach
to thermal equilibrium via a detailed balance condition.  Many of the
practical aspects of computer simulations, such as ``statistical errors''
and systematic errors due to the finite size of the simulated system or
the finite ``length'' of the simulated trajectory (or observation time,
respectively), are shared by all these simulation methods.

However, one important caveat needs to be made: it is quantum
mechanics that describes the basic physics of condensed matter,
and not classical mechanics. However, attempting a numerical
solution of the Schr\"odinger equation for a system of many nuclei
and electrons still is premature and not at all feasible even at
the fastest computers. Thus, one has to resort to approximations.
One very popular approach is the ``{\it ab initio} MD'' or
``Car-Parrinello-method'' \cite{12}, where one includes some
electronic degrees of freedom into MD via density functional
theory (DFT) \cite{13}. The huge advantage of this technique is
that one no longer relies on effective interatomic potentials,
which often are only phenomenologically chosen {\it ad hoc} assumptions,
lacking any firm quantum-chemical foundation. However, the
disadvantage of this technique is that it is several orders of magnitude
slower than classical MD, and hence only very short time scales
and very small systems are accessible. Furthermore, the method is
unsuitable to treat systems with van der Waals like forces, such as in rare
gases, where one still is better off with the simple Lennard-Jones
potential (perhaps amended by three-body forces) \cite{1}. Also,
normally ionic degrees of freedom are still treated classically.
Alternatively, one can still use effective potentials between ions
and/or neutral atoms  as in classical MD or MC, but rely on
quantum statistical mechanics for the ionic degrees of freedom:
this is achieved by path integral Monte Carlo (PIMC) \cite{14,15}
or path integral molecular dynamics (PIMD) \cite{16,17,18}. Such
techniques are indeed crucial for a study of solids at low
temperatures, to ensure that their thermal properties are
compatible with the third law of thermodynamics. For most fluids
(of course, quantum liquids such as $^3$He and $^4$He are an
exception) classical MD is fine, and will henceforth be considered
exclusively in this article.

What information do we then desire to extract from the
simulations? When we consider systems in thermal equilibrium, the
first task is to calculate static properties. E.g., in a fluid a
basic property is the static structure factor $S(k)$ \cite{19}

\begin{equation} \label{eq1}
S(k)= \langle | \delta \rho (\vec{k})|^2 \rangle ,
\end{equation}

\noindent
where $\delta \rho (\vec{k})$ is a spatial Fourier transform of
density fluctuations, $\vec{k}$ being the wavevector. In addition,
one wants to calculate time-dependent correlation functions, that
describe the decay of small thermal fluctuations with time. A
quantity of this type that will be discussed later is the
intermediate scattering function $S(k,t)$, 

\begin{equation} \label{eq2}
S(k,t)= \langle \delta \rho (-\vec{k}, 0) \,\delta \rho
(\vec{k},t) \rangle   .
\end{equation}

If one considers systems out of thermal equilibrium, an important
application of MD is the study of systems exhibiting a steady
state flow under shear deformation. The purpose of such NEMD
\cite{20,21} work can be the estimation of transport
coefficients (e.g.~the shear viscosity) if the deformation is
weak enough so that the system is still in the regime of linear
response \cite{19}. However, also the study of nonlinear phenomena
(such as ``shear thinning'', a decrease of the effective viscosity
with increasing shear rate \cite{20,21}) is of interest~\cite{berthier02}. In
addition, one can also study non-steady state transient behavior,
as it occurs e.g.~after a sudden quench from one state to another,
where one wishes to study the approach of the system to its new
thermal equilibrium. Classical examples of this problem are
nucleation of fluid droplets in a supersaturated gas or the
kinetics of phase separation in a binary mixture (``spinodal
decomposition'' \cite{22}). However, often such processes are too
slow, and cannot really be studied by NEMD, and instead one has to use 
simulations of coarse-grained models by Non-Equilibrium Monte Carlo
(NEMC) \cite{2,4,9,11}.

Now this list of simulation techniques and problems that one may
simulate sounds wonderful -- everything can be studied with
computer simulation methods, even problems outside the standard
realm of physics: the spontaneous formation of traffic jams on
highways \cite{23}, anomalous time-dependent autocorrelation
functions of heart beats of human beings suffering from heart
diseases \cite{24}, critical fluctuations at the stock market
before a crash \cite{25}, or shock waves that can destroy a silo used
in agriculture when the grains of corn stored in it flow out at
the wrong speed \cite{26}, etc. All these problems are in fact
simulated by theoretical physicists using techniques very similar
to MD or MC. However, one must nevertheless always keep in mind
that computer simulations -- as all other techniques! -- suffer
from some very important technical limitations that the
practitioner always must be aware of, just as an experimentalist
in an inelastic scattering experiment must be aware of that the
resolution of his energy and momentum transfer measurements limits
his data analysis, and that in addition there are statistical errors
due to limited intensity of the radiation source.

\subsection{Limitations of Computer Simulations:\\
 When to Choose Which Method?}

The main limitation of atomistic simulations comes from the fact
that one often must bridge many decades in spatial scale and even
more decades in time scale to describe the phenomena of interest
\cite{27,28}. As an example, we discuss here the problem of phase
separation of a mixture of a polymer (such as polybutadiene)
and its deuterated counterpart \cite{29}. If one carries out a
quenching experiment, suddenly reducing the temperature from a
high value in the one phase region to a lower value inside the
miscibility gap and records the equal-time structure factor $S(k)$,
Eq.~(\ref{eq1}), at different times $t$ after the quench, one
observes the growth of a peak at a position $k_m(t)$ \cite{29}.
Now polybutadiene is a flexible linear macromolecule, which in the
molten state forms random walk-like coils that exhibit nontrivial
structure from the scale of covalent C-C and C-H bonds (i.e., of
the order of 1~\AA) to the coil radius (which is of the order of
$10^2$~\AA, for the molecular weights used in the experiment). The
collective length scale $\ell(t) \approx 2 \pi/k_m(t)$ is of the
order of 1000~\AA~already in the initial stage of phase
separation, however. Clearly, such a study of cooperative
phenomena of a large number of chains would be prohibitively
difficult, if we would try a chemically realistic, atomistically
detailed description. Moreover, the description of effective
potentials driving this phase separation between protonated and
deuterated polymers which otherwise are chemically identical is
quite subtle: in the framework of classical statistical mechanics,
the masses of the particles cancel out from all static averages,
and hence such a mixture would be an ideal mixture, perfectly
miscible, no phase separation would ever occur. The reason that
phase separation is observed in the experiment \cite{29} is a
quantum effect, the zero point vibrational motions of hydrogens
and deuterons differ: the lighter hydrogens need more space and
this causes an effective repulsive interaction between unlike
species.

An even greater problem is the disparity of time scales encountered
in this example: while the time-scale of bond length and bond angle
vibrations is of the order of $10 ^{-13}$ s, already the time needed for a
coil to renew its configuration can be estimated as $10^{-5}$~s, 8 orders
of magnitude larger than the vibration times, for the considered molecular
weight \cite{29} and temperature of the experiment. The collective
dynamics, however, is even much slower, because the thermodynamic
driving forces are very weak. Thus, the experiment shows \cite{29} that
the interesting phenomena happen on the time scales from a few seconds
to 1000 seconds, when a scattering peak develops at $k_m(t)$ and first
grows more or less at the same position and then shifts to smaller and
smaller wavevectors as the coarsening of the structure proceeds. And
while for the case of phases separation in metallic alloys the situation
is better with respect to the length scale $k_m(t)$ \cite{22,30}, $k_m(t)
\sim 0.01 -0.1$~\AA$^{-1}$, the typical time scale is still from $0.1$~s
to $10^3$~s, and hence for a chemically realistic molecular dynamics
simulation, with a time step $\delta t$ in the range from 1 fs to 10 fs,
the task is quite hopeless, one would have to simulate over a range of
10$^{15}$ time steps which is many orders of magnitude more than what
is feasible nowadays.

Such slow phenomena as spinodal decomposition in solid metallic
alloys or fluid polymer mixtures can only be simulated by very
simplified coarse-grained models, where chemical detail is
sacrificed, and one applies non-equilibrium Monte Carlo (NEMC)
methods rather than MD. These coarse-grained simulations
nevertheless are very useful, both for solid alloys \cite{31} and
for polymers \cite{32}. In the latter case, several successive
chemical monomers along the chain are integrated into one
``effective bond''. Moreover, one also simulates relatively short
(unentangled) chains, and hence one does not attempt to study
chains of large molecular weight as done in the experiment
\cite{29}. Even with these simplifications, it is difficult to
deal with such long wavelength phenomena: the computer simulation
can never deal directly with the system in the thermodynamic
limit, one always can treat only a finite system. In this case of
spinodal decomposition of binary polymer mixtures \cite{32} a
cubic box containing a few hundred or a few thousand short polymer
chains is studied. To avoid surface effects at the boundaries of
the simulation box, periodic boundary conditions are used. Such
systems sometimes are called quasi-infinite, but this is not quite
true: the finite size of the system still has notable effects,
e.g.~the reciprocal space is ``quantized'', since the only wavevectors that are
compatible with the periodic boundary conditions are of the form

\begin{equation} \label{eq3}
\vec{k} = (k_x,k_y,k_z) = \frac{2\pi}{L} (n_x,n_y,n_z) \quad ,
\end{equation}

\noindent
where $L$ is the size of the box and $n_x,n_y,n_z$ are integers.

In practice one does find for this problem that the position
$k_m(t)$ where the peak grows occurs near values of $k$ where
$\mu < 10$, and hence the discreteness of $\vec{k}$--space is a
real practical problem. Despite such problems, the simulations of
collective phenomena \cite{31,32} are useful, but it is always
advisable to carefully pay attention to possible artifacts caused
by the finite size of the simulation box.

\subsection{Internal Dynamics of Polymer Chains: An Example on
What Can be Done When Models are Carefully Chosen}

This example of unmixing in polymer blends \cite{32} is not meant
to imply that MD simulations for polymers are not feasible at all:
on the contrary, MD work for polymers can be very useful and can
even be compared to experiment quantitatively, but only for
carefully selected problems! This is shown in an example
\cite{33,34} dealing with the relaxation of the configuration of
polymer coils. While for entangled chains with a degree of
polymerization of the order of $10^3$ the relaxation time $\tau_R$
of the coil configuration typically is at least $10^{-5}$~s,
the choice of shorter non-entangled chains such as C$_{100}$
H$_{202}$ brings $\tau_R$ down to about  $\tau_R
\approx 1$ ns, if a sufficiently high temperature is chosen ($T=509$ K
in this case). Fig.~\ref{fig1} shows a comparison of data for the
single-chain (normalized) coherent intermediate scattering function

\begin{equation} \label{eq4}
S(q,t) =\frac{1}{N_p} \sum\limits_{n,m=1}^{N_p} \langle \exp [i
\vec{q} \cdot (\vec{r}_n (t + t_0) - \vec{r}_m (t_0)] \rangle
\end{equation}

\noindent
obtained by the neutron spin echo technique, with MD simulation
results of a suitable model. In Eq.~(\ref{eq4}), $\vec{q}$ is the
scattering vector, $N_p$ is the degree of polymerization of the
chain ($N_p =100$ here), and $\vec{r}_m(t_0)$ is the position of
the $m$'th monomer of a chain ($1 \leq m \leq N_p)$ at time $t_0$.
The average $\langle \cdots \rangle$ in the simulation is taken
over all chains in the system and is also meant to represent the
time average over the time $t_0$. Note that we have already made
use of the fact that in thermal equilibrium a time-displaced
correlation function as written on the rhs of Eq.~(\ref{eq4})
depends only on the difference $t$ between the two times $t +
t_0$, $t_0$, and not on these two times separately.

It is seen from Fig.~\ref{fig1} that there is an almost perfect
agreement between the experimental data and the simulations, over
two decades in time and almost a decade in the wavevector, and in
this scaled plot there is no adjustable parameter whatsoever!
Thus, this agreement is by no means a trivial fitting exercise,
but rather it is truly significant. While the original conclusion
of the experimentalists was that their data prove the validity of
the Rouse model \cite{35,36}, one now knows, thanks to the
simulation (Fig.~\ref{fig2}), that this is not quite correct.
Remember that the Rouse model describes the polymer as a harmonic
chain, which experiences friction and stochastic forces which
represent the interactions with the surrounding chains. This
simplistic model contains only two parameters, an effective size
$\sigma$ of a bead, and the friction coefficient $\zeta$. Both are
known from independent estimates of other quantities in the
simulation (static structure factor $S(q,0)$ and self-diffusion
constant $D_R$ of the chains, respectively). So again one can
perform a comparison between the simulation and the theory without
any adjustable parameters whatsoever (Fig.~\ref{fig2}). One sees
that the Rouse model works very well for small values of
$q$, while it fails for larger wavelengths. Later experimental
work did confirm the conclusion of the simulations \cite{34} that
the Rouse model is indeed not so perfect as originally claimed.

This figure also shows that $S(q,t)$ for typical values of $q$
really does decay to zero on the scale of 1 to 10 nanoseconds. We
emphasize again that this was possible only due to carefully
chosen simulation parameters: C$_{100}$H$_{202}$ is
a relatively short chain, and $T=509$ K is a relatively high
temperature. The experiment deliberately was done for such
parameters to allow a meaningful comparison with a simulation. As
was said above, for other parameters it could easily happen that
the relaxation times for $q\approx R_g^{ -1}$ ($R_g$ being the
gyration radius of the chains) is larger by a factor of
$10^3$-$10^6$ than in the present case. While at large enough $q$
it still may be possible to study the relaxation in the ``time
window'' accessible for the scattering experiment, it would not
make sense to compare with a computer simulation: the latter
cannot reach thermal equilibrium if the Rouse time (the time
needed to equilibrate the configuration of the {\it whole} chain) is so
large. Note that often it is thought that there is always a direct
correspondence between quantities that one can study by inelastic
neutron scattering and the corresponding MD observations, since
both methods have a ``time window'' of about 1 ns. However, in an
experiment one can invest a time of days or even weeks to
carefully anneal the sample and thus prepare very good
equilibrium. But this is not the case in a simulation: in MD work one
almost always has to start from some initial state which is not
yet characteristic for the desired equilibrium, and let the system
relax towards equilibrium with a MD run that also can last only a
few nanoseconds! Therefore meaningful inelastic scattering
experiments can study the relaxation of fast degrees of freedom in
fluids rather close to the glass transition temperature of
polymers or other glass-forming systems, since the system is in
good thermal equilibrium. In contrast to this MD simulations of a corresponding model
can reach equilibrium only at much higher temperatures, however.
These simple considerations are almost obvious but nevertheless
often ignored -- therefore we emphasize these points here.

A related misconception is that a MD simulation is the better the
more chemical detail is included in the model: however, this
attitude is completely wrong! In fact, the level of detail in the
simulations of Paul {\it et al.}~\cite{33,34} normally did neither
include fluctuations in the lengths of the C-C bonds along the
chain backbone, nor were the hydrogen atoms explicitly included:
the bond length was kept at the experimental value
$\ell_{\rm cc}= 1.53$~\AA, and the CH$_2$ groups were treated
as ``united atoms'' (also no full distinction between the CH$_3$
groups at the chain ends and the CH$_2$ groups in the interior was
made). If one includes the hydrogens explicitly, the program
is about a factor 10 slower, and a modest gain in accuracy
of the model is more than outweighed by about 3 times larger
statistical errors. Note that this dramatic slowing down of the
code is inevitable due to the very stiff potentials that need to
be included in such an ``all atom''--calculation and which
necessitate then a particularly small time step (and also the number of
atoms is 3 times larger). The potentials actually used for
the simulations shown in Figs.~\ref{fig1} and~\ref{fig2} are bond
angle potentials $U(\Theta)$ and torsion  potentials $U(\phi)$ of
the form

\begin{equation} \label{eq5}
U(\Theta) = \frac{1}{2} k_\Theta (\cos \Theta - \cos \Theta_0)^2
\, , \ \ \ U(\phi) = \sum\limits_{n=0}^{5} \,\, a_n \, \cos^n(\phi)
\, ,
\end{equation}

\noindent
while non-bonded monomers interact with a Lennard-Jones potential

\begin{equation} \label{eq6}
U(r_{ij} )= 4 \varepsilon_{\alpha \beta} [(\sigma / r_{ij})^{12} -
(\sigma /r_{ij})^6] \quad , \quad \alpha, \beta \in \{\mbox{CH}_2,
\mbox{CH}_3\} \quad .
\end{equation}

\noindent
The parameters $k_\Theta$, $\Theta_0$, $a_n$, $\varepsilon_{\alpha
\beta}$, $\sigma$ are given in \cite{33,34}. In principle, such
effective potentials for classical MD simulations should be
deduced from quantum-chemical calculations. However, polyethylene
is a much too large molecule to do that: only the bond angle and
torsional potentials have a quantum chemical basis, although even
on this issue there is no full consensus between different groups.
Of course, it is clear that the Lennard-Jones potential used here,
\{Eq.~(\ref{eq6})\} is completely {\it ad hoc}, and the parameters are
just optimized in order to fit as many experimental data as
possible. For the case of C$_{100}$H$_{202}$, a box
of linear dimension $L=50$~\AA ~allows to include 40 chains in the
simulation, and finite size effects are then negligible for the
wavevectors of interest. It is
clear that a treatment of longer chains would not only require
larger box sizes but also much longer runs, and therefore are very
difficult. We also emphasize that Eq.~(\ref{eq6}) is not a good
basis to describe interactions between chemically dissimilar
polymers with sufficient accuracy.

\section{MD algorithms and some simulation ``knowhow''}
\subsection{Classical Mechanics and the Verlet Algorithm}

We consider a system of $N$ particles (atoms) with Cartesian
coordinates $\vec{X}=\{\vec{r}_i\}$, $i=1, \cdots, N$, in a
$d$-dimensional space. The dynamics then is described by Newton's
equations of motion,

\begin{equation} \label{eq7}
m_i \ddot{\vec{r}}_i=- \frac{\partial U_{\rm pot}}{\partial\
\vec{r}_i}= \vec{f}_i \, ,
\end{equation}

\noindent
$m_i$ being the mass of the $i$'th particle, and $\vec{f}_i$ the
force acting on it, which we assume to be entirely due to
interactions with other particles. Thus, the potential
$U_{\rm pot} (\vec{X})= U_{\rm pot} (\vec{r}_1, \cdots,
\vec{r}_N)$ is written as

\begin{equation} \label{eq8}
U_{\rm pot}=\sum\limits_{i=1}^{N-1} \, \sum\limits_{j>i}^{N} \, U
(\vec{r}_{ij} )\quad , \quad \vec{r}_{ij} = \vec{r}_i - \vec{r}_j
\, \, ,
\end{equation}

\noindent
where in the last step we have furthermore made the simplifying
assumption that $U_{\rm pot}$ is pairwise additive (this
assumption, which is reasonable for simple liquids, is not really
necessary, it can be generalized whenever appropriate, we only
make it for the sake of simplicity of our presentation). Thus

\begin{equation} \label{eq9}
\vec{f}_i=-\sum\limits_{j( \neq i)} \, \partial U (r_{ij}) /
\partial \vec{r}_i = \sum\limits _{j (\neq i)} \, \vec{f}_{ij}
\quad .
\end{equation}

\noindent
The total energy

\begin{equation} \label{eq10}
E=E_{\rm kin} + U_{\rm pot}=\sum\limits_{i=1}^{N} \,
\frac{1}{2} m_i {\dot{\vec{r}}}_i\,^2 + U_{\rm pot}
\end{equation}

\noindent
is a constant of motion,

\begin{equation} \label{eq11}
\frac{dE} {dt} = \sum\limits_{i=1}^N m_i {\dot{\vec{r}}}_i \,
{\ddot{\vec{r}}}_i - \sum\limits_{i=1}^N \, {\dot{\vec{r}}}_i
\cdot \vec{f}_i =0 \quad .
\end{equation}

MD now means that Newton's equations of motion are integrated
numerically. A computationally efficient scheme is the so-called Verlet
algorithm~\cite{37}. To derive it, we expand $\vec{r}_i(t)$ forward and
backward in time,

\begin{equation} \label{eq12}
\vec{r}_i (t + \delta t)= \vec{r}_i(t) + \delta t \vec{v}_i
(t) + \frac{1}{2 m_i} (\delta t)^2 \vec{f}_i (t) + \frac{1}{6}
(\delta t)^3 \vec{b}_i(t) + O ((\delta t)^4),
\end{equation}

\begin{equation} \label{eq13}
\vec{r}_i (t - \delta t)= \vec{r}_i (t) - \delta t
\vec{v}_i (t) + \frac{1} {2 m_i} (\delta t)^2 \vec{f}_i (t)
- \frac{1}{6} (\delta t)^3 \vec{b}_i (t) + O (( \delta t)^4) \, .
\end{equation}

\noindent
Here $\vec{v}_i(t)$ is the velocity of the $i$'th particle at
time $t$, and $\vec{b}_i(t)$ is the corresponding vector that
appears in the third order of the Taylor expansion.

For ${\ddot{\vec{r}}} (t)$ we have already substituted Newton's
equation. Adding both expansions, the odd terms cancel, and hence

\begin{equation} \label{eq14}
\vec{r}_i(t + \delta t)= 2 \vec{r}_i (t) - \vec{r}_i (t - \delta
t) + \frac{1} {m_i} (\delta t)^2 \vec{f}_i (t) + 0 ((\delta t)^4)
.
\end{equation}

\noindent
Subtraction of the expansions yields the velocity

\begin{equation} \label{eq15}
\vec{v}_i (t) = \frac{1}{2 (\delta t)} \, [ \vec{r}_i (t +
\delta t )- \vec{r}_i (t - \delta t)] + 0 ((\delta t)^3) \quad.
\end{equation}

\noindent
This Verlet algorithm [1,7,19,37,38] is manifestly time
reversible: exchange of $\vec{r}_i (t + \delta t)$ and
$\vec{r}_i(t - \delta t)$ yields the propagator for the time
evolution going backward in time, and changes the sign of the
velocity.

A peculiarity of the Verlet algorithm is the fact that the
updating of the velocities is one step behind: the velocities at
time $t$ can only be calculated, see Eq.~(\ref{eq15}), after the
positions at time $t + \delta t$ have become available. The
updating of positions and velocities is synchronized in the
``Velocity Verlet Algorithm'' \cite{39}. Using

\begin{equation} \label{16}
\vec{r}_i ( t + \delta t) = \vec{r}_i (t) + \delta t
\vec{v}_i + \frac{1}{2 m_i} (\delta t)^2 \vec{f}_i (t)
\end{equation}

\noindent
together with the corresponding time-reversed equation

\begin{equation} \label{eq17}
\vec{r}_i (t) =\vec{r}_i (t + \delta t) - \delta t
\vec{v}_i (t + \delta t) + \frac{1} {2 m_i}(\delta t)^2
\vec{f} _i (t + \delta t)
\end{equation}

\noindent
one obtains by adding these equations

\begin{equation} \label{eq18}
\vec{v}_i (t + \delta t)= \vec{v}_i (t) + \frac{\delta t}{2 m_i} 
[\vec{f}_i (t) + \vec{f}_i (t + \delta t)] \, .
\end{equation}

\noindent
The propagation of the velocities hence requires that both the
forces of the present and the future configurations are known. The
second set of forces can be determined as soon as the coordinates at
time $t + \delta t$ have been obtained. Note that both the Verlet
and the Velocity Verlet Algorithms are completely equivalent: they
produce identical trajectories.

At this point it is of interest to ask what is the scale for the
time step. For simplicity we consider liquid argon, the
``fruit fly'' for MD. The argon atoms are assumed to interact with
a Lennard-Jones potential as considered in Eq.~(\ref{eq6}), with
parameters $\sigma \approx 3.4$~\AA, $\varepsilon/k_B\approx 120$
K,  and $m \approx 6.6 \times 10^{-23}$g. Rescaling coordinates by
writing $\vec{r}\,^* \equiv \vec{r}/\sigma$, we obtain

\begin{equation} \label{eq19}
\vec{r}\,^*(t + \delta t) = 2 \vec{r}\,^* (t) - \vec{r}\,^* (t
-\delta t) - (\delta t)^2 \,\, \frac{48 \varepsilon}{m \sigma^2}
\,\, \frac{\vec{r}\,^*_{ij}}{| \vec{r}\,^*_{ij}|} \,
\sum\limits_{j(\neq i)} \, [(r^*_{ij})^{-13} - (r\,^*_{ij})^{-7}
/2] .
\end{equation}

\noindent
From this equation we see that the natural time unit $\tau_0$ is

\begin{equation} \label{eq20}
\tau_0=\left(\frac{m \sigma^2}{48 \varepsilon}\right)^{1/2} ,
\end{equation}

\noindent
which is roughly $\tau_0 \approx3.1 \times 10^{-13}$ s for Argon.
In order to keep numerical integration errors small, we need to choose
the time step $\delta t$ so small that the third term on the rhs of
Eq.~(\ref{eq19}) is significantly smaller that the first and second
term. For the Lennard-Jones system this usually means $\delta t^*=0.03$,
which translates into real time units $\delta t \approx 10^{-14}$
s for the time step. So even with a million time steps (rescaled time
$t^*=t/\tau_0=30000)$ one only reaches a real time of about 10 ns.

It is also useful to go beyond this ``pedestrian approach'' to classical
mechanics, discussing more formally the time evolution in phase space,
combining the Cartesian coordinates of all the particles $(\vec{X})$
and their momenta $\vec{P}=(\vec{p}_1, \vec{p}_2, \cdots , \vec{p}_N)$
into a point in a 2$Nd$-dimensional space, $\Gamma \equiv (\vec{X},
\vec{P})$. Liouville's theorem says that the flow in phase space has the
property that it is incompressible, phase space volume is conserved. One
can formally write for any observable $A$

\begin{equation} \label{eq21}
A(\Gamma,t)= \widehat{U}(t) A (\Gamma,0) ,
\end{equation}

\noindent
where the propagator $\widehat{U}(t)$ is a unitary operator,

\begin{equation} \label{eq22}
\widehat{U}(t)=\exp (i \, {\mathcal{\widehat{L}}}\, t) ,
\end{equation}

\noindent
${\mathcal{\widehat{L}}}$ being the Liouville operator. Now the
Verlet algorithm has the desirable feature that (to order ($\delta
t)^3)$ unitarity is preserved. This fact may to some extent
explain why the Verlet algorithm is such a particularly stable
algorithm over very long times (for a detailed discussion of the
stability of the Verlet algorithm see Ref.~\cite{tuckerman}).

\subsection{Estimation of Intensive Thermodynamic Variables; the
$NVT$ Ensemble}

If the energy $E$ were rigorously conserved by the algorithm it would
realize states distributed according to the microcanonical ($NVE$)
ensemble of statistical mechanics. Of course due to integration errors
the energy is not strictly conserved and hence one has to make sure that
the chosen time step is small enough, as will be discussed below -- let us
disregard this problem for the moment and assume that the microcanonical
ensemble is realized perfectly.  How do we then estimate the thermodynamic
variables of interest, such as temperature $T$ and pressure $p$?

One can make use of the fact that in classical statistical
mechanics the distributions of positions and velocities factorize;
one simply has a Maxwell-Boltzmann distribution for the velocities and
thus $T$ can be inferred from the kinetic energy of the particles.
Defining a ``kinetic temperature'' $\mathcal{T}$ as follows (for
$d=3$)

\begin{equation} \label{eq23}
\mathcal{T}= \frac{2}{3 k_BN} \, E_{\rm kin} =\frac{1} {3 k_B
N} \, \sum\limits_{i=1}^{N} \, m_i \vec{v}\,^2_i \quad ,
\end{equation}

\noindent
the desired estimate of the temperature of the system is then 
computed as the average $T= \langle \mathcal{T} \rangle$. It must
be emphasized that for finite $N$ in the microcanonical ensemble
there occur fluctuations in temperature, described by \cite{40}

\begin{equation} \label{eq24}
\frac {[\langle {\mathcal{T}^2}\rangle - \langle
{\mathcal{T}}\rangle ^2]}{\langle {\mathcal{T}}\rangle ^2} =
\frac{2} {3N} \left(1-\frac{3k_B}{2C_v}\right) ,
\end{equation}

\noindent
where $C_v$ is the specific heat of the system (at constant
volume). Considering that one always works with a finite length of
the MD run, these fluctuations cause a statistical error in the
estimation of the temperature.

The pressure $p$ can be estimated using the virial Theorem. The
dynamical variable whose average yields the pressure is defined as

\begin{equation} \label{eq25}
\mathcal{P} = \frac{1}{3V} \left(2E_{\rm kin}+ \sum ^{3N}_{i=1} \vec{r}_i
\cdot \vec{f}_i \right) \; ,
\end{equation}

\noindent
such that

\begin{equation} \label{eq26}
p= \langle \mathcal{P} \rangle = \frac{Nk_BT}{V} + \frac {1}{3V}
\sum^N _{i=1} \left< \vec{r}_i \cdot \vec{f}_i \right> \; .
\end{equation}

Now we return to the problem that the energy is not strictly
conserved by the algorithm described so far. The original recipe
used in early MD work was to rescale the velocities from time to
time such that one corrects for the drift in the energy.
Alternatively, one can draw randomly velocities from time to time
from a Maxwell Boltzmann distribution. Both methods have the
disadvantage that dynamic correlations are somewhat disturbed, and
one realizes in this way neither a microcanonical nor a canonical
ensemble. In the canonical ($NVT$) ensemble, temperature is a
given variable and therefore not fluctuating at all, and instead
of the fluctuations of $\mathcal{T}$ \{Eq.~(\ref{eq24})\} one
encounters fluctuations of the total energy $E$, described by
another fluctuation relation: 

\begin{equation} \label{eq27}
N C_v/k_B = (1/k_BT)^2(\langle {\mathcal{H}}^2 \rangle _{_{NVT}} -
\langle {\mathcal{H}} \rangle ^2_{_{NVT}})\; .
\end{equation}

\noindent
Here $\mathcal{H}$ is the Hamiltonian of the system. 
With methods in which one rescales the velocities, both
temperature and energy are fluctuating, and neither
Eq.~(\ref{eq24}) nor Eq.~(\ref{eq27}) hold.

However, one can introduce a MD method that does reproduce the
canonical ($NVT$) ensemble exactly: one has to extend the
Lagrangian of the system by a variable, representing the
thermostat which has a fictitious mass $Q$. This yields the
Nos\'e-Hoover algorithm \cite{41,42}. Newton's equations of motion
are then extended by a ``friction'' term,

\begin{equation}\label{eq28}
\ddot{\vec{r}}_i = \vec{f}_i/m_i - \zeta \dot{\vec{r}}_i \; .
\end{equation}

\noindent
The friction coefficient $\zeta(t)$ fluctuates in time around
zero, according to the equation

\begin{equation} \label{eq29}
\dot{\zeta} =\frac{1} {Q} \, \left[\sum\limits_{i=1}^{N} m_i
\vec{v}_i \,^2 - 3 Nk_BT \right] \quad .
\end{equation}

\noindent
Thus, $\zeta(t)$ responds to the imbalance between the
instantaneous kinetic energy and the intended canonical average
(remember $\sum\limits_i m_i \langle \vec{v}\,^2_i \rangle
= 3 Nk_BT).$ Although the total energy is no longer conserved, one
can identify a conserved energy-like quantity $\mathcal{H}'$,

\begin{equation} \label{eq30}
\mathcal{H}' = \mathcal{H} + \frac{1}{2} Q \zeta^2 + 3 Nk_BT \int
\zeta (t') d t' \quad ,
\end{equation}

\noindent
for which one can show that $d \mathcal{H}'/dt=0$ .

Also in this case it is true that dynamical correlation functions
between observables such as $\langle A(0) A (t)\rangle $ are not
precisely identical to the microcanonical ones. However, in many
cases of practical interest (e.g., dense systems of flexible
bead-spring-type chain molecules representing a glass-forming
polymer melt \cite{43}) the difference between the result for
$\langle A(0) A(t) \rangle$ from a strictly microcanonical run and
a run using this Nos\'e-Hoover thermostat is negligibly small
\cite{43}.

It is also possible to realize the isothermal-isobaric ($NpT$) ensemble,
where the pressure is given and rather the volume fluctuates, by
coupling to the so-called ``Andersen barostat'' \cite{44}. We shall not
describe this here, but refer the reader to the literature \cite{1,7,38}
for details.

\subsection{Diffusion, Hydrodynamic Slowing Down, and the Einstein
Relation}

As a final point of this discussion of technical aspects, let us
discuss how one extracts diffusion constants and other transport
coefficients from MD simulations. We consider first the
phenomenological description of diffusion in a single-component
system in terms of Fick's law since this will allow us to discuss
concepts such as hydrodynamic slowing down, the Einstein relation
for the diffusion coefficient $D$ in terms of mean square
displacements of particles, and the Green-Kubo formula providing a
link with the velocity autocorrelation function \cite{1,19,45}.

Fick's law states that there is a current of particles caused if there
is a (particle) density gradient (or a gradient in the concentration)
and this current acts to reduce the gradient. For small and slow density
variations this is a linear relation,

\begin{equation} \label{eq31}
\vec{j} (\vec{r}, t) =- D \nabla \rho (\vec{r}, t) ,
\end{equation}

\noindent
where $D$ is the diffusion constant. If we combine this
(phenomenological!) constitutive equation of irreversible
thermodynamics with the (exact!) continuity equation which expresses
the fact that the total particle number $N$ is conserved,

\begin{equation} \label{eq32}
\partial \rho(\vec{r}, t) / \partial t + \nabla \cdot \vec{j}(\vec{r}, t)
=0 \quad,
\end{equation}

\noindent
we get the well-known diffusion equation,

\begin{equation} \label{eq33}
\partial \rho(\vec{r}, t) / \partial t = D \nabla^2 \rho (\vec{r},
t) \quad.
\end{equation}

This equation is easily solved both in real space and in
reciprocal space. In reciprocal space a simple exponential
relaxation of the Fourier components of the density fluctuations
$\delta \rho_{\vec{k}} (t)$ results,

\begin{equation} \label{eq34}
\delta \rho (\vec{r}, t) = \rho (\vec{r}, t) - \langle \rho\rangle
= \int \delta \rho({\vec{k}},t) \exp (i \, \vec{k} \cdot \vec{r})
d \vec{k} \quad  ,
\end{equation}

\noindent
since

\begin{equation} \label{eq35}
\frac{d} {dt} \, \delta \rho(\vec{k},t) = - D k^2 \delta
\rho(\vec{k},t) \quad  , \quad \delta \rho(\vec{k},t)=
\delta \rho(\vec{k},0) \exp (-D k^2 t).
\end{equation}

\noindent
We recognize that the associate relaxation time $\tau_{\vec{k}}$
diverges as $k \rightarrow 0$,

\begin{equation} \label{eq36}
\tau_{\vec{k}} = (D k^2 )^{ -1} .
\end{equation}

\noindent
Therefore we see that the dynamic correlation function of density
fluctuations for long wavelengths in a diffusive system decays very
slowly,

\begin{equation} \label{eq37}
S(\vec{k}, t) \equiv \langle \delta \rho(-\vec{k},0) \delta
\rho(\vec{k},t) \rangle = S (k) \exp (- D k^2 t) \quad ,
\end{equation}

\noindent
where $S(k)$ is the static structure factor from Eq.~(\ref{eq1}):
$S(k)= \langle \delta \rho(-\vec{k},0) \delta \rho(\vec{k},0) \rangle $.
Eqs.~(\ref{eq36}),~(\ref{eq37}) demonstrate the so-called ``hydrodynamic
slowing down'' \cite{46}. On the one hand, analysis of the intermediate
scattering function as given by 
Eq.~(\ref{eq37}) allows to extract the diffusion constant.
On the other hand, this hydrodynamic
slowing down is a difficulty for equilibration in the $NVT$ ensemble,
{\it both} for MD and for MC simulations. Remember that for the
simulations of fluids, it is often convenient to use as an initial state
a regular arrangement of the particles in the box, by simply putting them
on the sites of the crystal lattice. This implies that $S(k \rightarrow
0)=0$ in the initial state. If the simulated volume is large, so that
the smallest wavevector $k_{\rm min}= 2\pi/L$ is small, it will
take a very long time until $S (k_{\rm min})$ reaches its equilibrium
value. In judging whether or not full equilibrium has been achieved,
one hence cannot rely on the recipe advocated in some of the early
simulation literature to check whether the internal energy has reached
its equilibrium value. This was okay for the early work, where only
64 or 256 particles were simulated, and hence $k_{\rm min}$ was not
small. Being interested in the simulation of much larger systems today,
the issue of equilibrating long wavelength fluctuations properly needs
careful consideration \cite{47}.

It is also useful to examine the diffusion equation in real space
since it readily allows to derive the Einstein relation for the
mean square displacement of a diffusing particle. Let us take as
an initial condition for the solution of Eq.~(\ref{eq33}) a 
$\delta-$function, $\rho (\vec{r}, t = 0) = \delta (\vec{r} -\vec{r}_0)$
\{note that the normalization $\int \rho (\vec{r}, t) d \vec{r}
=1$ means that $\rho (\vec{r}, t)$ can be interpreted physically
as the conditional probability to find a particle at $\vec{r}$
after a time $t$ provided this particle was at $\vec{r}=\vec{r}_0$
at time $t=0$\}. Now the solution of Eq.~(\ref{eq33}) is simply a
Gaussian distribution,

\begin{equation} \label{eq38}
\rho (\vec{r}, t) = ( 4 \pi D t ) ^{-d/2} \, \exp
[-(\vec{r}-\vec{r}_0)^2 / (4 D t) ] \quad ,
\end{equation}

\noindent
$d$ denoting once more the dimension of space. The squared half
width of this distribution increases linear with time $t$, and so
does the mean square displacement:

\begin{equation} \label{eq39}
\langle [\delta \vec{r} (t) ]^2 \rangle =\langle [ \vec{r} (t) -
\vec{r}_0]^2 \rangle= 2 d D t \quad , \quad t \rightarrow \infty .
\end{equation}

\noindent
We have added here the restriction to consider large times
($t\rightarrow \infty$) since then this Einstein relation,
Eq.~(\ref{eq39}), holds quite generally, while typically the
simple diffusion equation, Eq.~(\ref{eq33}), will not hold on
small length scales (of the order of a few atomic diameters or less) and
the corresponding short times. This fact will be evident with the
examples discussed later. In fact, Eq.~(\ref{eq39}) is routinely
used in simulations to compute the (self-)diffusion constants.

\subsection{Green Kubo Relations and Transport Coefficients}

It is also interesting to note that there is a relation between
the self-diffusion constant and the velocity autocorrelation
function of a diffusing particle \cite{19}.

Writing

\begin{equation} \label{eq40}
\vec{r} (t) -\vec{r}_0 = \int\limits_0^t \vec{v} (t') d t'
\quad ,
\end{equation}

\noindent
the mean square displacement can be expressed as follows:

\begin{equation} \label{eq41}
\langle [\vec{r} (t)-\vec{r}_0]^2 \rangle = \int\limits_0^t d t'
\int\limits^t_0 d t'' \langle \vec{v} (t'') \cdot
\vec{v} (t') \rangle= d \int\limits_0^t d t'
\int\limits_0^t d t'' Z (t'' - t'),
\end{equation}

\noindent
with $Z(t''-t') \equiv \langle v_\alpha (t'')
v_\alpha (t') \rangle$. In this notation, we imply that in
equilibrium translation invariance holds, so $Z$ depends only on
the time difference $t''-t'$, not on the two times $t', t''$
separately. $v_\alpha $ is one of the $d$ Cartesian components of
$\vec{v}$. Rearranging the domain of integration in
Eq.~(\ref{eq41}) readily yields

\begin{equation} \label{eq42}
\langle [\vec{r}(t)-\vec{r}_0]^2\rangle = 2 dt \int\limits_0^t
(1-s/t) Z (s) ds \underset{t \rightarrow \infty}{\longrightarrow}
2dDt \; ,
\end{equation}

\noindent
with

\begin{equation} \label{eq43}
D = \int\limits^\infty _ 0 Z(s)ds = \int\limits _0 ^\infty \langle
v_ \alpha (0) v_\alpha (t) \rangle dt .
\end{equation}

\noindent
This result that the self-diffusion constant is the time integral of the
velocity autocorrelation function is a special case of a ``Green-Kubo
relation'' \cite{19}. Another example of such a relation is that the
shear viscosity $\eta$ is related to time correlations of the off-diagonal
components of the pressure tensor $\sigma _{xy}$,

\begin{equation} \label{eq44}
\eta = \frac{1}{Vk_BT} \int\limits^{\infty}_0 dt \langle \sigma _{xy} (0)
\sigma _{xy}(t) \rangle \; ,
\end{equation}
where
\begin{equation} \label{eq45}
\sigma _{xy} = \sum ^N_{i=1} \left\{ m_i v_i^x v_i^y + \frac {1}{2}
\sum_{j( \neq i)} x_{ij} f_y (r_{ij}) \right\} \; ,
\end{equation}

\noindent
$f_y$ being the $y$-component of the force with which particles
$i,j$ interact, $f_y=- \partial U(r_{ij}) / \partial y$, cf.
Eq.~(\ref{eq9}), $\vec{r}_{ij}=\vec{r}_i-\vec{r}_j$. Similarly,
for the thermal conductivity $\lambda_T$ we need to consider
correlations of the energy current density \cite{19,45}

\begin{equation} \label{eq46}
\lambda_T =\frac{1} {Vk_B T^2} \, \int\limits_0^\infty dt \langle
j_z^e (0) j_z^e (t) \rangle \quad ,
\end{equation}

\begin{equation} \label{eq47}
j_z^e = \frac{d} {dt} \big[\sum\limits_{i=1}^N \, z_i (\frac{1}{2}
m_i \vec{v}_i\,^2 + \sum\limits_{j (\neq i)} \, U
(r_{ij}))\big] \quad.
\end{equation}

Finally, the electrical conductivity $\sigma_{\rm el}$ can be
obtained from correlations of the current density $J_x^{\rm el}
$ of the electrical charges $q_i$ \cite{19,45}

\begin{equation} \label{eq48}
\sigma_{\rm el}= \frac{1} {Vk_BT} \, \int\limits_0^\infty \, d
t \langle J_x^{\rm el} (0)
J_x^{\rm el} (t) \rangle \quad ,
\end{equation}

\begin{equation} \label{eq49}
J_x^{\rm el} = \sum\limits_{i=1}^N \, q_i
v_i^x \quad .
\end{equation}

\noindent
All these relations are in fact useful for MD simulations.

\section{Application to the Example of molten Silica}

As an example for the type of results that we can get using the
Einstein and Green-Kubo relations that were just discussed, we show
in Fig.~\ref{fig3} an Arrhenius plot for the viscosity of molten
SiO$_2$ \cite{48}, and compare the MD results with experimental data
\cite{49}. ``Arrhenius plot'' means, the logarithm of the viscosity is
plotted vs.~inverse temperature, since then an Arrhenius law [$\eta(T)
\propto \exp [E_{\rm A}/(k_BT)]$, where $E_{\rm A}$ is an ``activation
energy''] shows up as a straight line in the plot, with a slope $E_{\rm
A}$.  The experimental data included here \cite{49} is indeed compatible
with such a law. However, the extremely large values of the viscosity
mean that the relaxation time of this melt is in the microsecond or even
millisecond range. Therefore the MD results cannot be obtained in the
same temperature range where the experimental data were taken, but only
at considerably higher temperatures, where experiments are no longer
possible. Nevertheless, these simulation results (together with results
for many other correlation functions \cite{48} that will not be shown
here) do have a great theoretical relevance, since they show that even
SiO$_2$ has a regime of temperatures where the so-called mode coupling
theory of the glass transition \cite{50} can be applied. Since this
regime of temperatures is inaccessible with laboratory experiments, the
hypothesis was advanced that one needs several classes for glass-forming
fluids with a fundamentally different dynamical behavior. Fig.~\ref{fig3}
shows that ``computer experiments'' can complement laboratory experiments
by providing results for physical quantities in a parameter range which
can not be studied in the laboratory. Another interesting aspect of the
data shown in Fig.~\ref{fig3} is that the Stokes-Einstein relation,
linking the temperature dependence of the viscosity to that of the
self-diffusion constant(s), does not hold.

Simulating real glass-forming fluids such as SiO$_2$ is difficult
for various reasons: a great problem is again the choice of a
suitable force field. In Fig.~\ref{fig3}, the so-called
BKS-potential \cite{51} was used, which has been proposed on the
basis of quantum-chemical calculations. It contains Coulomb-like
interactions, but with effective charges $q_i \{ i \in $ Si, O \}
rather than the true ionic charges, and a short range Buckingham
potential,

\begin{equation} \label{eq50}
U(r_{ij}) = \frac{q_i q_j e^2}{r_{ij}} + A_{ij} \,\exp (- B_{ij}
r_{ij}) - C_{ij} /r_{ij}^6 \quad ,
\end{equation}

\noindent
where $e$ is the elementary charge, $q_0=-1.2$, $q_{\rm Si} =+2.4$,
and the constants $A_{ij}$, $B_{ij}$ and $C_{ij}$ can be found in the
original reference \cite{51}. It is somewhat surprising that a clever
choice of these phenomenological constants allows to describe the
directional covalent bonding (typically a Si atom is in the center of a
tetrahedron, with the oxygens at the corners), although one uses just
pair potentials. Nevertheless, the simulations are still technically
very difficult: the long range Coulomb interaction necessitates the use
of the time-consuming Ewald summation techniques \cite{1,2,3}; the scale
for the potential is in the eV energy range and varies rather rapidly
with distance. Therefore one needs to use a rather small MD time step,
namely $\delta t=1.6$ fs.

If one produces amorphous glassy structures by a MD simulation with
such a model, the most plausible procedure is the same as the one used
in the glass factory. One starts at a very high temperature $T_0$,
and then one cools the system gradually down at constant pressure,
so the temperature $T(t)$ varies with time, e.g. $T(t)=T_0- \gamma t$,
$\gamma$ being the cooling rate. While the structures obtained in this
way look qualitatively reasonable \cite{52}, one must be aware of the
fact that the cooling rates applied in the simulation are extremely large
$(\gamma=10^{12}$~K/s or even higher), which means that they are at least
a factor $10^{12}$ larger than those used in the experiment. Thus, one
must expect that the quantitative details of the results (including also
the density of the material) will depend on the cooling rate distinctly,
and this is what actually is found \cite{52}.

Better results are obtained if one proceeds in a slightly different way,
where one no longer tries to predict the density of the model system
from the simulation, but rather fixes it to the experimental value. The
system then is carefully equilibrated at a ``moderately high'' temperature
before one cools it down to the desired temperature. ``Moderately high''
for pure SiO$_2$ means $T=2750$ K -- then equilibration requires a
run which is already 20 ns long. To avoid finite size effects, a box
of around 48~\AA~linear dimension needs to be taken, containing about
8000 atoms. This calculation requires substantial CPU resources (it was
done \cite{48} at a CRAY-T3E parallel supercomputer, performing force
parallelization) so it is presently not easy to do much better.

In any case, with this procedure one does obtain the static structure
factor of silica glass at room temperature in very good agreement with
experiment. This can be inferred, e.g, from Fig.~\ref{fig4} where we show
the static structure factor as measured in a neutron-scattering experiment
and which is the weighted sum of the three partial static structure
factors~\cite{19,48,49}. Since there are no adjustable parameters
whatsoever involved in this comparison, this agreement is significant.

Fig.~\ref{fig5} shows now an Arrhenius plot for the self-diffusion
constants, using scales such that the experimental data \cite{54,55}
can be included together with the simulation results.  From this graph
it becomes obvious that in principle one would like that the simulation
spans over 16 decades in dynamics, a task clearly impossible for MD. So
there is again a gap between the range where simulations in thermal
equilibrium are possible and the temperature range where the experimental
data can be taken. But it is gratifying that the simulation can predict
the activation energies for the self-diffusion constants almost correctly.

Now the real strength of the MD simulations is that one can record
in a single calculation a great variety of different static and
dynamic properties simultaneously. For instances, one can study
the time dependence of autocorrelation functions of the
generalized temperature fluctuation $\delta T_{\vec{q}}(t)$, which
is the Fourier transform of the local ``temperature'' $T(\vec{r},
t)$,

\begin{equation} \label{eq51}
T(\vec{r}, t) = \sum\limits_{i=1}^N \, \frac{\vec{p}_i^2}{3 k_B m_i} \;
\delta (\vec{r} -\vec{r}_i (t)) \quad .
\end{equation}

\noindent
One sees (Fig.~\ref{fig6}a) \cite{56} that for large wavelength $q$
this correlator is rapidly relaxing but again one notes a slowing down at
small $q$. This is another example for ``hydrodynamic slowing down''. Of
course, the considerable statistical scatter of the results shown in
Fig.~\ref{fig6}a is somewhat disturbing. But it has to be mentioned
that these results are already based on averages over 100-200 runs,
corresponding to an effort of several years if one would run the problem
on a single CPU, and hence it is currently difficult to do better.

Defining now from the time-displaced correlation function,

\begin{equation}
\Phi_{TT} (q,t)= \langle \delta T_{-\vec{q}} (0) \delta
T_{\vec{q}} (t) \rangle \  , 
\end{equation}

\noindent
a relaxation time $\tau$ by the condition

\begin{equation} \label{eq52}
\Phi_{TT} (q, t= \tau) \equiv \Phi_{TT} (q,t=0)/10
\quad ,
\end{equation}

\noindent
one obtains the data shown in Fig.~\ref{fig6}b. Here a plot of
$\tau q^2$ vs. $q^2 $ is presented, since theory predicts
\cite{56}

\begin{equation} \label{eq53}
\tau=\frac{\rho C_p}{\lambda_T} \,\, \frac{1} {q^2} +
{\rm const} \quad ,
\end{equation}

\noindent
where $\rho=N/V$ is the density, $C_p$ the heat capacity, and $\lambda_T$
is the thermal conductivity.  Since $\rho$ is given and $C_p$ can be
estimated independently, one obtains a rough estimate of the thermal
conductivity $\lambda_T$ from these data (Fig.~\ref{fig6}b). This
estimate is in reasonable qualitative agreement with experiments, which
indicate that for $T \geq 1000$ K $\lambda_T$ is only weakly dependent
on temperature and in the range $2 \leq \lambda_T \leq$ 3 W/Km \cite{56}.

As a last example, the temperature dependence of the longitudinal
$(c_{\ell})$ and the transverse $(c_t)$ sound velocity is shown in
Fig.~\ref{fig7} \cite{57}. These results were obtained from an analysis
of the corresponding time-displaced current-current correlation functions
which were Fourier transformed into frequency space. Undamped propagation
of sound manifests itself by $\delta-$functions $\delta (\omega -
\omega_q$) with $\omega_q=c_{l,t} q$ for $q \rightarrow 0$ in these
correlators \cite{57}. Of course, in a liquid for $q \rightarrow 0$ no
static shear can be maintained and hence in that case only $c_{\ell}$
is well-defined.  However, for not too small $q$ both longitudinal and
transverse correlators show broad peaks, and the positions of these peaks
are shown in Fig.~\ref{fig7}. Again one notes very nice agreement with
corresponding experimental data \cite{58}.

\section{A Brief Introduction to Non-Equilibrium Molecular
Dynamics (NEMD)}

Already in Eq.~(\ref{eq31}), we have considered the situation that
the density in the system may deviate from its constant
equilibrium value, and we have postulated a linear relation
between the particle density current and the gradient of the local
density $\nabla \rho (\vec{r},t)$. Similarly, we also can assume
that there is no longer complete thermal equilibrium, as far as
the temperature is concerned, but only ``local equilibrium'': we
may still assume a Maxwell-Boltzmann velocity distribution of the
particles but with a temperature $T(\vec{r}, t)$ that slowly
varies in space. Thus, a current of energy, i.e.~heat, is created,
the coefficient between the temperature gradient and the energy
current density being the thermal conductivity $\lambda_T$,
yielding Fourier's law of heat conduction,

\begin{equation} \label{eq54}
\vec{j}_Q = - \lambda _T [\nabla T(\vec{r}, t)] .
\end{equation}

Now in reality the situation is not so simple, since energy
density and particle number density are coupled variables. Each
gradient therefore produces also a current of the other variable.
So we have to generalize the set of flow equations
Eq.~(\ref{eq31}),~(\ref{eq54}) to a matrix form, involving the
Onsager coefficients $\Lambda_{\alpha \beta}$,

\begin{equation}  \label{eq55}
\vec{j}_Q= \Lambda_{QQ} \nabla \left( \frac{1}{k_BT} \right) - 
\Lambda_{Qi} \nabla \left( \frac{\mu_i}{k_BT} \right) \quad ,
\end{equation}

\begin{equation} \label{eq56}
\vec{j}_i= \Lambda_{iQ} \nabla \left( \frac{1}{k_BT} \right) -
\Lambda_{ii} \nabla \left( \frac{\mu_i}{k_BT} \right) \quad,
\end{equation}

\noindent
with $\mu_i$ being the chemical potential of particle species $i$
(the generalization of Eqs.~(\ref{eq55}), ~(\ref{eq56}) to systems
containing several different species of particles then is
obvious). Also, we have written the set of constitutive equations
entirely in terms of gradients of intensive variables, rather than
using the density of an extensive variable as in Eq.~(\ref{eq31}).
Of course, using thermodynamics, the corresponding relation is
easily established, since with Eq.~(\ref{eq31}) on obtains:

\begin{equation} \label{eq57}
\frac{\nabla \mu_i}{k_BT} = \left( \frac{\partial \mu_i}{\partial \rho_i} \right)_{T} \,
\frac{1}{k_BT} \, \nabla \rho_i \,\,  \Longrightarrow D_i =
\frac{\Lambda_{ii}}{k_BT} 
\left( \frac{\partial \mu_i}{\partial \rho_i} \right)_{T} \quad .
\end{equation}

In NEMD simulations one usually sets up a stationary gradient by
suitable boundary conditions creating a stationary current through
the system. In the case of a flow of particles, this clearly is
compatible with periodic boundary conditions: particles leaving
the simulation box at the right simultaneously reenter at the
left. For instance, we may confine a fluid between two parallel
plates and let a constant force act on the particles
\cite{20,21,59,60,61} but schemes where one avoids external walls
altogether also are possible \cite{20,21,62}. In the latter case,
one can obtain a homogeneous flow, while in the former case the
velocity profile is inhomogeneous, and also the structure of the
fluid close to the external walls is modified.

Now an important aspect of all such stationary flows is that
entropy is produced: the entropy production per unit time is

\begin{eqnarray} \label{eq58}
\frac{ds}{dt} = \Lambda_{QQ} \left[ \nabla \frac{1}{k_BT} \right]^2 + 
(\Lambda_{Qi} + \Lambda_{iQ} ) 
\left[ \left( \nabla \frac{1}{k_BT} \right)
\left( \frac{\nabla \mu_i}{k_BT} \right) \right] \nonumber\\
+ \Lambda_{ii} \left[ \frac{\nabla \mu_i}{k_BT} \right]^2 \quad .
\end{eqnarray}

\noindent
Although this entropy production per unit time is small for small
gradients, nevertheless the heat that is produced makes a strictly
microcanonical simulation impossible: the system would heat up
steadily. Thus one always has to use a thermostat.

As an example for such applications of NEMD, we now briefly review
some results \cite{60} obtained for a bead-spring model of short
polymer chains confined between two parallel plates. The
interaction between the beads is a simple Lennard-Jones (LJ)
potential  of the form as written in Eq.~(\ref{eq6}) \{truncated
at $r_c= 2.24 \sigma$ and shifted so that $U(r=r_c)=0$, as usual\}. The
``spring potential'' between neighboring effective monomeric units
is taken to be the finitely extensible nonlinear elastic (FENE)
potential,

\begin{equation} \label{eq59}
U_{\rm FENE} (\ell) = - 15 R_0^2 \ln [1 - (\ell/R_o)^2] \,\,
\end{equation}

\noindent
choosing ``Lennard-Jones units'' for length, energy and
temperature as usual, $\sigma =1$, $\varepsilon=1$, $k_B=1$ \{and
also the mass of the effective monomers $m=1$, which fixes the
unit of time as well, cf. Eq.~(\ref{eq20})\}. Choosing $R_0=1.5$
one finds that the total bond potential (LJ + FENE) has its
minimum at $\ell_0 \approx 0.96$, while the LJ potential has its
minimum at $r_0=2^{1/6} \approx1.13$. This competition between
these two length scales prevents crystallization of this melt, and
hence is the physical reason that this model is a very good model
for a glass-forming polymer melt \cite{43,63}. Note that unlike the
atomistic model \cite{33,34} discussed in Sec. 1.3 of the present
article, neither a bond angle potential nor a torsional potential
is included in this case. In fact, one envisages that the ``effective
monomers'' are formed by combining a few subsequent chemical
monomers along the backbone of the chain molecule into one such
effective unit of a ``coarse-grained'' model \cite{8,9,27}.

For a chain length $N=10$ a box of linear dimensions $L_x \times L_y
\times D$, with $L_x = L_y=10.05$, $D=20$, containing 200 chains in the
system is a good choice~\cite{60}.
We used periodic boundary conditions in $x$ and $y$
directions, while in the remaining $z$-direction one places walls
formed by atoms on a triangular lattice, at equilibrium positions
$\vec{r}_{i,{\rm eq}} $ with $z_{i, {\rm eq}}=\pm D/2$.
These wall atoms are bound to their equilibrium positions with a
harmonic potential, $U_{\rm harm}(\vec{r}_i)=\frac{1}{2} K_h
(\vec{r}_i-\vec{r}_{i, {\rm eq}})^2$, with $K_h=100$.

In order to represent the physical effect of the atoms in the
(massive) walls other than those in the first layers of the wall,
one coarse-grains them into an effectively repulsive background
potential $U_{\rm wall}(z) = \varepsilon(\sigma/z)^9$, where
$z\equiv|z_{\rm mon}-z_{\rm wall}|$, $z_{\rm mon}$
being the $z$-coordinate of the considered monomer, and
$z_{\rm wall}=\pm (\sigma + D/2)$ the effective positions of
the second layer on top of the walls. In order to create a flow in
this geometry, a force $\vec{F}^e$ acts in the $+x$-direction on
each monomer, and one also needs to specify the interaction
between the monomers and the atoms in the top layers of the walls,
which determine the ``hydrodynamic boundary conditions'' of the
resulting Poisseuille flow \cite{20,59,60,61}.

In Ref. \cite{60} this interaction was chosen of the same LJ form
as the interaction between non-bonded effective monomers, but with
different parameters, $\sigma_{\rm wm}=2^{-1/6}$,
$\varepsilon_{\rm wm}=2$. In this way one creates a ``stick''
boundary condition for the flow, while for
$\sigma_{\rm wm}=1$, $\varepsilon_{\rm wm}=1$ one would
have a strong partial slip, and the estimation of transport
coefficients from the velocity and temperature profiles would be
more difficult.

Fig.~\ref{fig8} shows typical results for the density and velocity profile
across the slit. One can recognize a very pronounced ``layering''
phenomenon near both walls, i.e.~$\rho(z)$ exhibits very strong
oscillations, but these oscillations are only weakly dependent on
temperature, and furthermore in the center region of the thin polymer
film ($-5 \leq z\leq +5$) the density profile is essentially flat,
exhibiting the behavior of the bulk. Using the radial distribution
function $g(\vec{r})$, with a vector $\vec{r}$ parallel to the walls,
it has been checked that the system has still a fluid like structure
(and not crystallized, for instance) near the walls where the strong
layering phenomena occurs.

The central region $(-5\leq z \leq +5$) then is used to fit the
Poisseuille-type flow profile (well-known from hydrodynamics) to
the data, i.e.~\cite{64}

\begin{equation} \label{eq60}
u_x(z)=-\rho_0 F^e (z^2-z^2_{\rm wall}- 2 z_{\rm wall}
\,\delta)/ \eta \quad,
\end{equation}

\noindent
where $\rho_0= \rho \,(z=0)$ is the bulk density, and $\delta$ has
the physical interpretation of a slip length \{note $\partial
u_x(z)/\partial z {\underset{z=z_{\rm wall}} {\mid}}
=u_x(z=z_{\rm wall}) / \delta\}.$ Fig.~\ref{fig8}b shows
that excellent fits to Eq.~(\ref{eq60}) are in fact obtained,
thus yielding an estimate of the viscosity $\eta$.

At this point a comment on how the thermostating is done is in
order. The most natural way (used in the laboratory experiments)
would be to enforce isothermal conditions through the walls,
keeping them at constant temperature by coupling the wall atoms to
a thermostat. The heat created in the center of the polymer film
through the viscous flow then leads to a stationary temperature
profile, and along the resulting temperature gradients the heat is
transported towards the walls. Of course, if the viscous flow is
too fast (when the force amplitude $F^e$ is too large) one enters
a regime of very nonlinear response, with strong temperature
variations and large gradients, and then descriptions based on
linear irreversible thermodynamics
\{Eqs.~(\ref{eq31}),~(\ref{eq51})--(\ref{eq60})\} are not really
applicable in a strict sense. As an example, Fig.~\ref{fig9} shows
temperature profiles across the film observed in a simulation of
this type. One sees that for $F^e \geq 0.1$ the temperature in the
center of the film is strongly enhanced, and one needs to choose
$F^e \leq 0.05$ to stay within the linear response regime. Of
course, the study of nonlinear phenomena far from equilibrium
(such as ``shear thinning'' \cite{21}, the decrease of the
effective viscosity $\eta_{\rm eff}$ with increasing shear
rate) may be of interest in itself, but this is out of
consideration here.

Fig.~\ref{fig9} also demonstrates that the temperature profiles
can also be fitted to the theoretical profiles resulting from the
solutions of the phenomenological hydrodynamic equations
\cite{64,65}

\begin{equation} \label{eq61}
T(z)= T_{\rm wall} + \frac{(\rho_0F^e D^2)^2}{192 \lambda _T
\eta} \left[ 1- \left(\frac{2z}{D}\right)^4 \right]\; .
\end{equation}

\noindent
In the temperature range of Fig.~\ref{fig9}, one obtains $3 \leq
\lambda _T \leq 4.4$ as a result for the thermal conductivity. We
also note that the data shown in Fig.~\ref{fig9} do not give any
evidence for an additional term not included in Fourier's law,
Eq.~(\ref{eq54}), namely a strain rate coupling that
would lead to an additional quadratic term $\{\propto -
(2z/D)^2\}$ in Eq.~(\ref{eq61}) \cite{59}.

Since in a glass forming fluid the viscosity is strongly
temperature dependent, already a small variation of $T(z)$ across
the film creates problems for the use of Eq. (\ref{eq61}) at low
temperature. It thus is preferable to apply the thermostating
algorithm not only to the wall atoms, but also to the fluid atoms,
noting that in the presence of the flow the temperature is defined
as $T=m \langle (\vec{v}-\langle \vec{v} \rangle) ^2 \rangle/(3k_B)$
instead of $T= \langle {\mathcal{T}} \rangle$
\{Eq.(~\ref{eq23})\}. Since here $\langle \vec{v} \rangle =
(u_x(z),0,0)$, one has during the equilibration to determine the
profile $u_x(z)$ self-consistently \cite{60}.

Fig.~\ref{fig10} shows the result for the shear viscosity $\eta$
and the lateral self-diffusion constant $D_\bot$ (obtained from
mean square displacements in the $y$-direction perpendicular to
the flow). One sees that one can obtain rather precise results for
$\eta$ and monitor the increase of $\eta$ over about two decades.
Both $\eta$ and $D_ \bot$ can be fitted well by the 
Vogel-Fulcher-Tammann (VFT) \cite{63} relation,

\begin{equation} \label{eq62}
\eta (T) = \eta (\infty) \exp [E_\eta / (T-T_0)] \; , \quad D_\bot
= D (\infty) \exp [-E_D /(T-T_0)] \; ,
\end{equation}

\noindent
where the VFT temperature $T_0$ is the same for both quantities,
while the effective activation energies $E_\eta , E_D$ are
slightly different. This difference is consistent with a direct
examination of the ratio $\ell = k_BT/(4\pi D_\bot \eta)$,
which should be a characteristic constant length if the
Stokes-Einstein relation holds (cf. Fig.~\ref{fig3}): similarly as
for SiO$_2$, one finds a systematic decrease of this ratio as the
temperature gets lower \cite{60}.

\section{Concluding Comments}
\begin{figure} 
\psfig{file=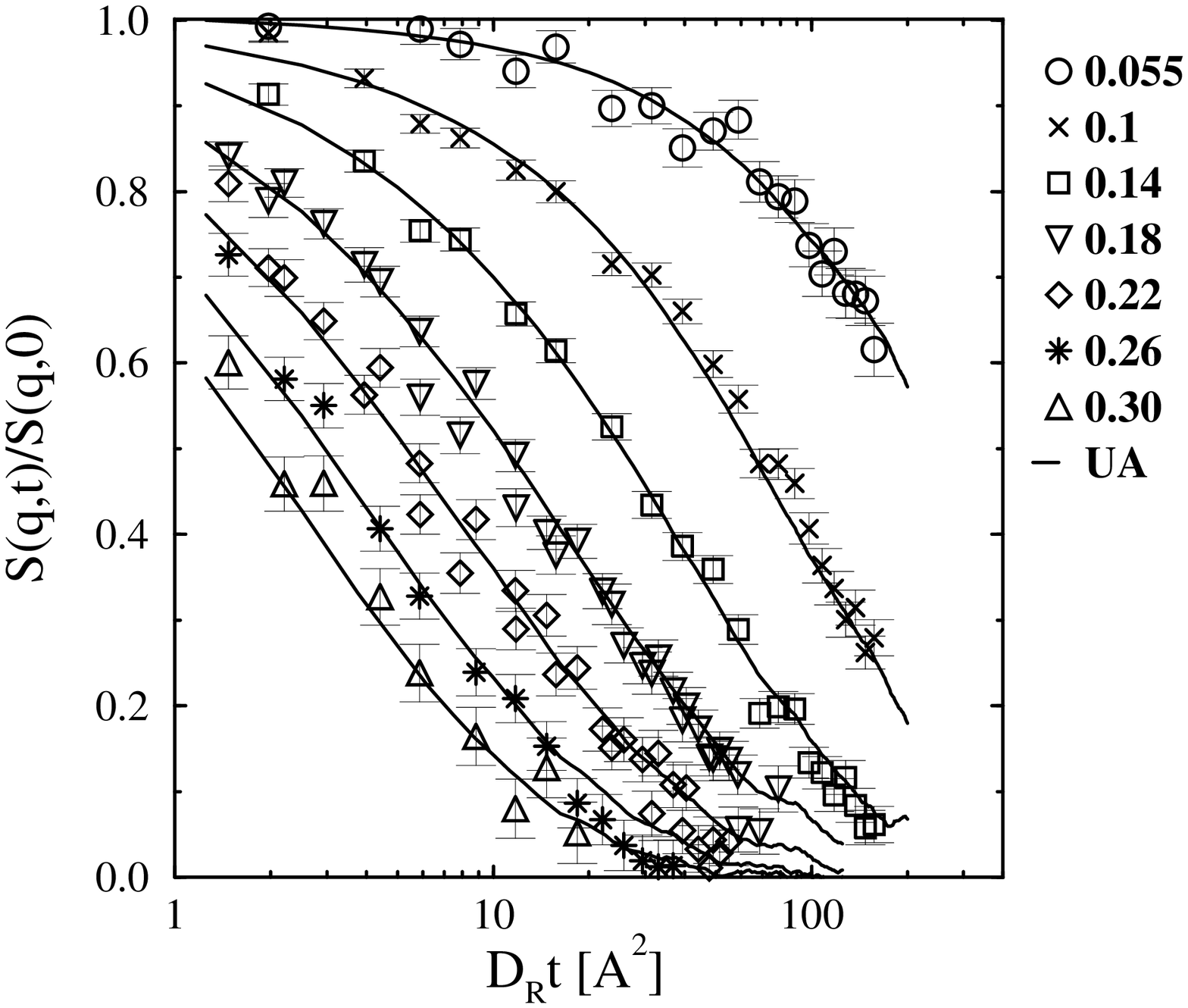,height=12cm}
\vspace*{-1.cm}
\caption{\label{fig1} Intermediate dynamic
structure factor versus scaled time as obtained from a neutron
spin echo experiment (symbols) and computer simulation results
(full curves). The time axis is scaled by the respective center of
mass diffusion coefficients $D_R$. The ordinate axis is scaled by
the static (equal-time) single chain structure factor $S(q,0)$.
The different symbols represent the different values of $q$, in
units of~\AA$^{-1}$, as explained in the legend. From Paul
\cite{34}.} \end{figure}

\begin{figure} 
\psfig{file=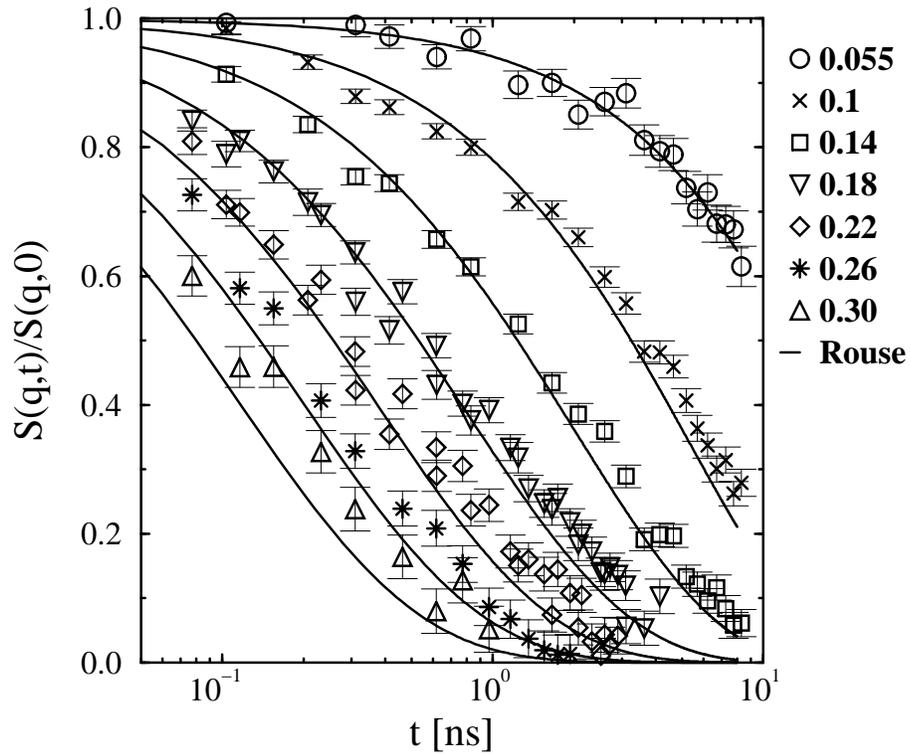,height=12cm}
\vspace*{-1.cm}
\caption{\label{fig2} Normalized coherent intermediate 
scattering function for C$_{100}$H$_{202}$ plotted
versus time, for the wavelength $q$ (in~\AA$^{-1})$ as indicated
in the legend. The full curves are the Rouse model prediction, the
symbols denote the simulation results. From Paul \cite{34}.}
\end{figure}

\begin{figure} 
\psfig{file=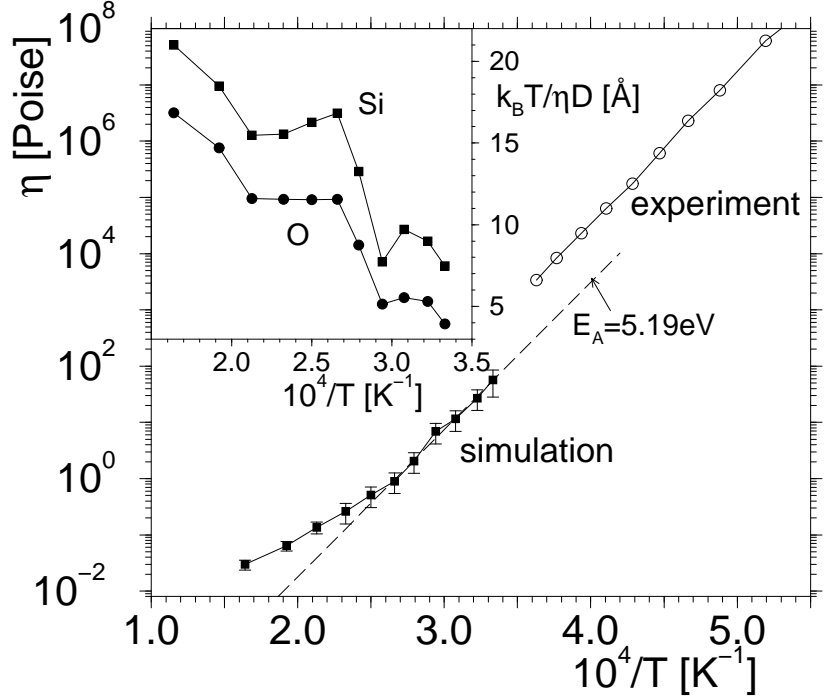,height=9cm}
\caption{\label{fig3}Molecular Dynamics results (filled squares) for
the viscosity of the BKS model for molten SiO$_2$ plotted vs.
inverse temperature. The dashed straight line indicates an
Arrhenius fit with an activation energy $E_{\rm A}=5.19$~eV.
The experimental data \cite{49} (open circles) are compatible with this value but
would suggest a slightly larger preexponential factor. Note that
for SiO$_2$ the analysis of other correlation functions at very
high temperature suggests a critical temperature $T_{\rm c} = 3330$~K of
mode coupling theory, therefore $\eta(T >T_{\rm c})$ deviates from the
Arrhenius law. The inset shows that the Stokes-Einstein relation,
$k_BT/(\eta D)=$~const, where $D$ are the Si or O
self-diffusion constants, does not hold in the regime of
temperatures studied. From Horbach and Kob \cite{48}.}
\end{figure}

\begin{figure} 
\psfig{file=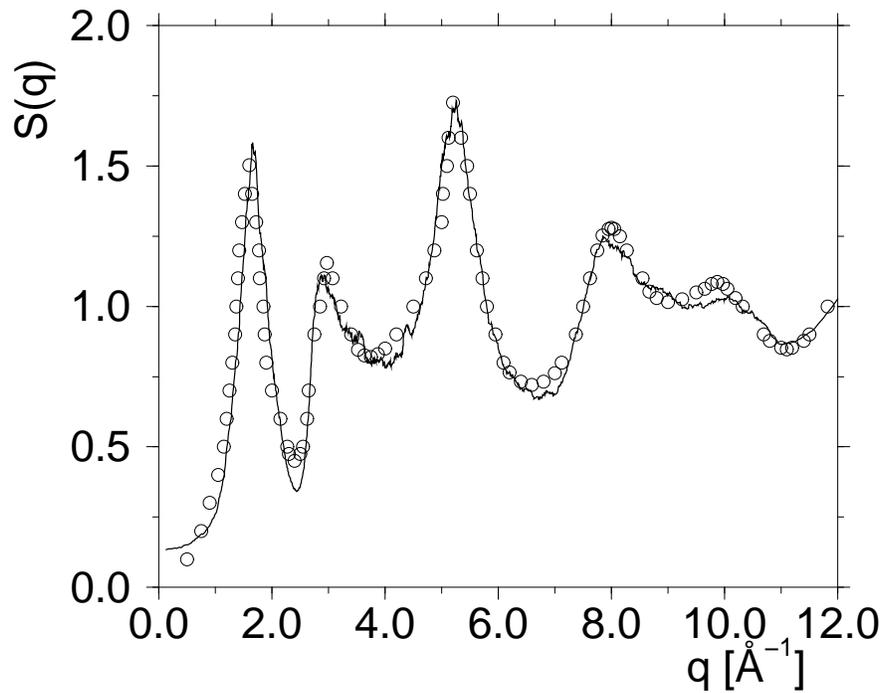,height=9cm}
\caption{\label{fig4}Static neutron structure
factor of SiO$_2$ at room temperature $(T=300$~K), plotted versus
wavelength $q$. The full curve is the MD simulation \cite{48},
using the experimental scattering lengths for Si and O atoms,
while the symbols are the neutron scattering data of Price and
Carpenter \cite{53}. From Horbach and Kob \cite{48}.}
\end{figure}

\begin{figure} 
\psfig{file=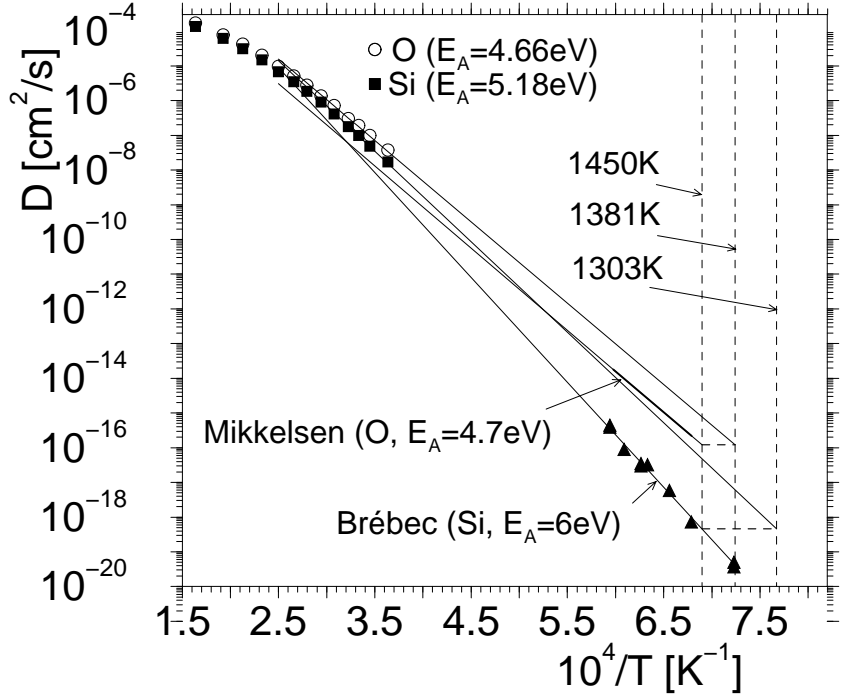,height=9cm}
\caption{\label{fig5} Plot of the self-diffusion
constant $D$ of silicon atoms (Si) and oxygen atoms (O) in molten
SiO$_2$ as a function of inverse temperature. The symbols in the
upper left part are the results from MD simulations and the data
in the lower right part stems from experiments \cite{54,55}. The
thin straight lines show simple Arrhenius behavior ($D \propto
\exp [-E_{\rm A}/(k_BT)]$) with an activation
energy $E_{\rm A}$, as indicated in the figure. The vertical
broken lines indicate the experimental glass transition
temperatures, $T_g=1450$ K, as well as values for $T_g$ that one
obtains if one extrapolates the data from the simulations to low
temperatures and then estimates $T_{{\rm g}}$ from the experimental
value of the O diffusion constant
($D_{\rm O}(T=T_{\rm g}^{\rm sim})=10^{-16}$ cm$^2$/s\;$\Longrightarrow
T_{\rm g}^{\rm sim}=1381$ K ) or the Si diffusion constant,
respectively ($D_{\rm Si} (T=T_{\rm g}^{\rm sim})=5.10^{-19}$
cm$^2$/s \,$\Longrightarrow T_{\rm g}^{\rm sim}=1303$ K). From
Horbach and Kob \cite{48}.}
\end{figure}

\begin{figure}
\psfig{file=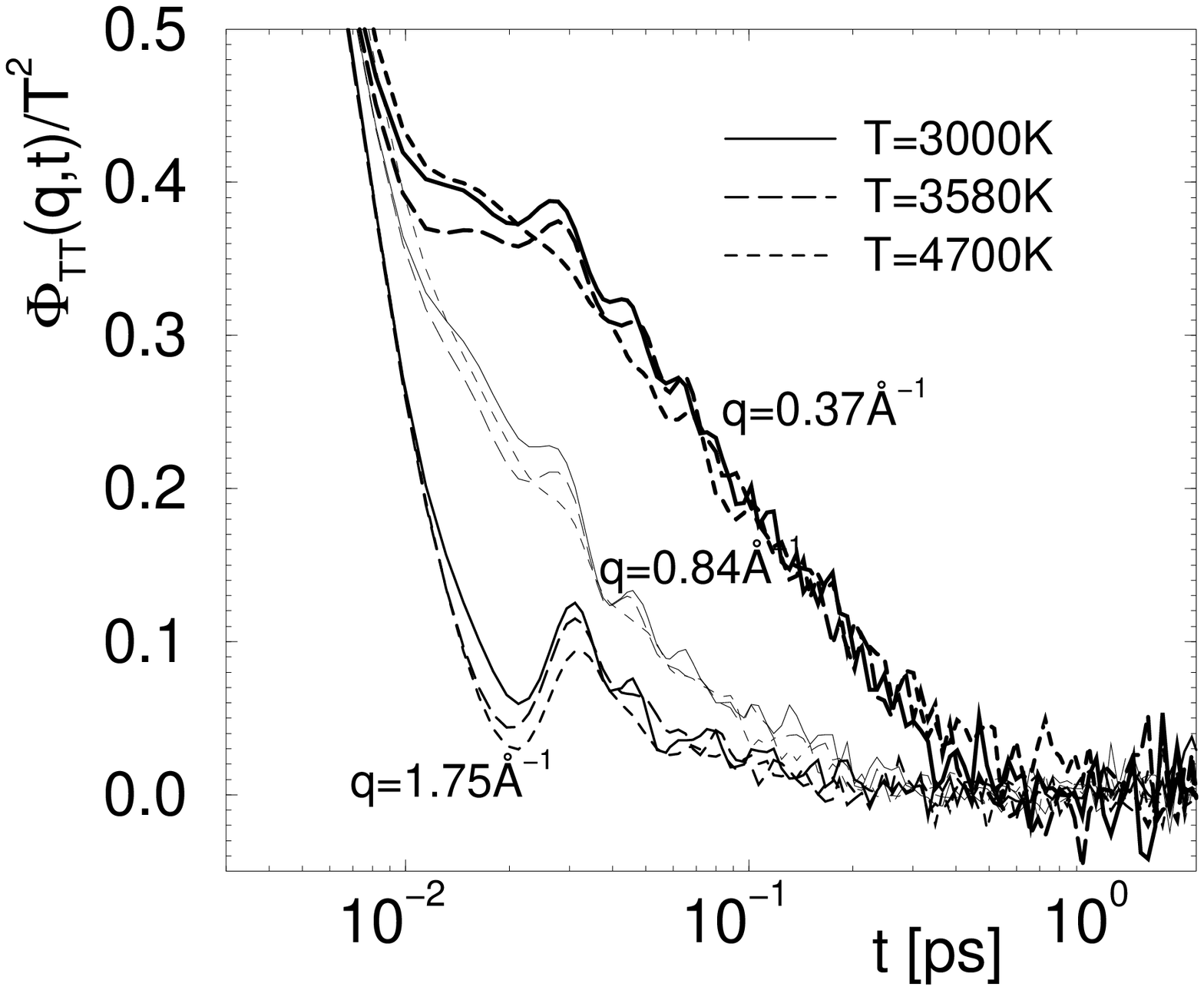,height=7.5cm}
\psfig{file=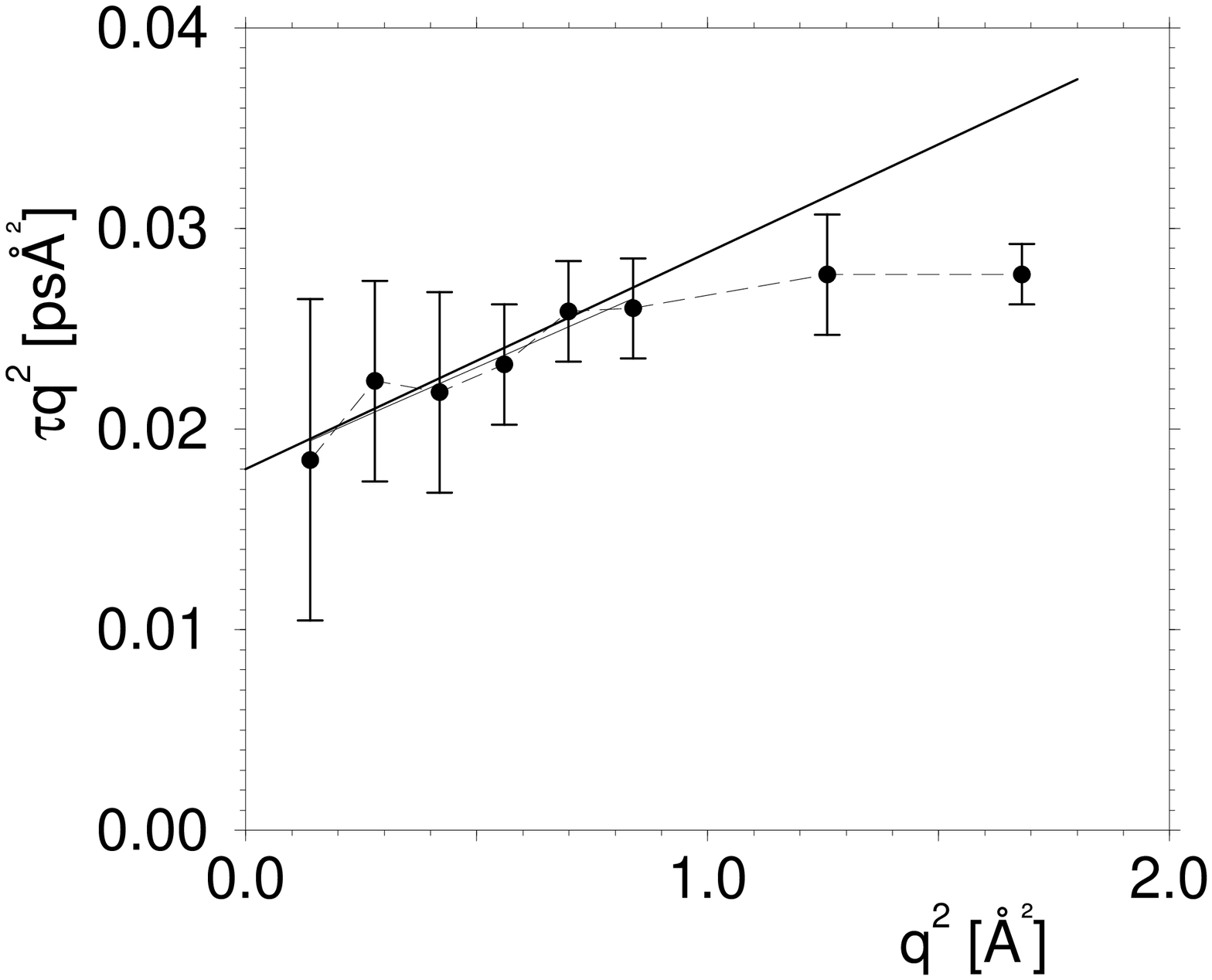,height=7.5cm}
\caption{\label{fig6}
a) Time-dependence of the autocorrelation function
$\Phi_{TT}(q,t)/T^2$ of the generalized temperature
fluctuation $\delta T_q(t)$ for SiO$_2$, showing data for three
temperatures $T$ and three choices of wavevector $q$ as indicated.
All data refer to systems containing 224 oxygen and 112 silicon
atoms. b) Plot of wavevector squared times the relaxation time
$\tau$, as defined in Eq.~(\ref{eq52}), versus $q^2$. From the
extrapolated intercept for $q^2 \rightarrow 0$, the estimate
$\lambda_T \approx 2.4$ W/(Km) for the thermal conductivity
$\lambda_T$ is extracted. From Scheidler {\it et al.}~\cite{56}.}
\end{figure}

\begin{figure} 
\psfig{file=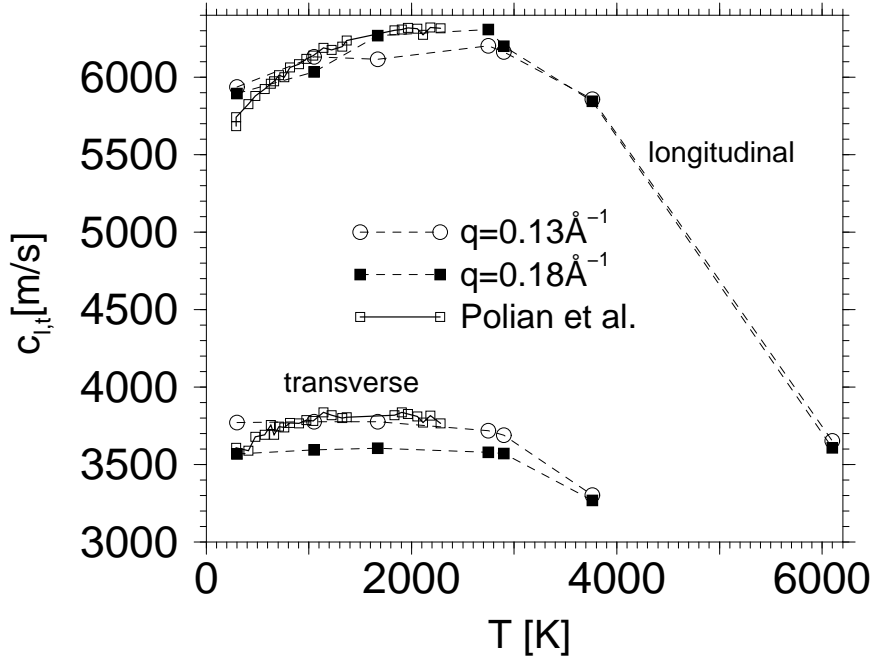,height=8.5cm}
\caption{\label{fig7}
Longitudinal ($c_\ell$) and transverse ($c_t$) sound velocities of SiO$_2$
plotted vs.~temperature. These quantities were determined from the
frequency positions of the maxima of the corresponding
longitudinal and transverse current correlation functions at
$q=0.13$~\AA$^{-1}$ (open circles) and $q=0.18$~\AA$^{-1}$ (filled
squares). Also included are the experimental data of Polian {\it et
al.}~\cite{58} which are multiplied with the factor
$(2.2/2.37)^{1/2}$ since the simulation was done at a density of
$\rho_{\rm sim}=2.37$~g/cm$^2$ while the experiment was done
for $\rho_{\rm exp}=2.2$~g/cm$^2$. From Horbach {\it et al.}~\cite{57}.}
\end{figure}

\begin{figure}
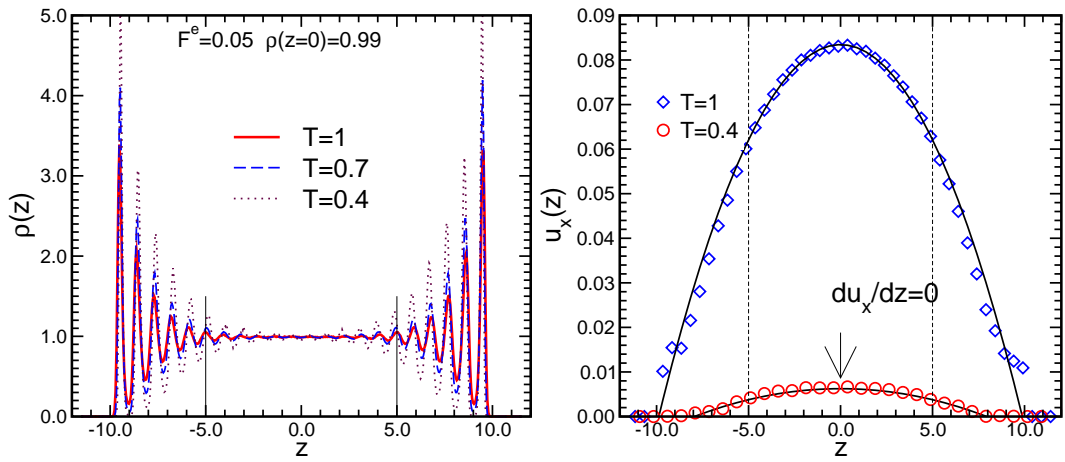
 
\hspace*{-1.5cm}
\psfig{file=./fig8a.eps,height=6cm}
\psfig{file=./fig8b.eps,height=6cm}
\caption{\label{fig8}
a) Monomer density profiles of a polymer melt confined in a slit. The bulk
density is $\rho=\rho(z=0)=0.99$ and the amplitude of the force in
the $x$-direction is $F^e=0.05$. Three temperatures are included as
indicated. Note that the coordinate origin for the $z$-axis is
chosen in the center of the polymer film. b) Velocity profile
$u_x(z)$ for two temperatures (as indicated) and otherwise the
same conditions as in panel a). The parabolic curves indicate the
fitted Poisseuille flow profiles, Eq.~(\ref{eq60}). From Varnik
and Binder \cite{60}.}
\end{figure}

\begin{figure} 
\psfig{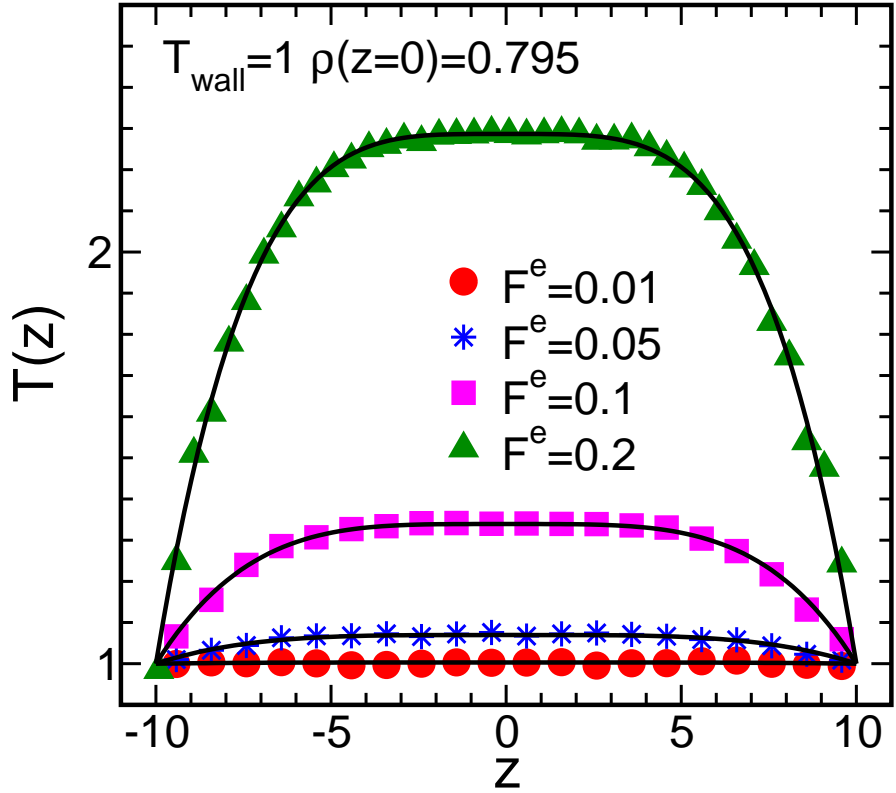}
\caption{\label{fig9}
A comparison of the temperature profiles resulting from NEMD
simulations (symbols) with the theoretical prediction
Eq.~(\ref{eq61}) (lines) for various values of $F^e$ at a wall
temperature $T_{\rm wall}=1$ and a bulk density
$\rho(z=0)=0.795$. Contrary to the results shown in Figs.~\ref{fig8} and \ref{fig10}, 
here the fluid particles were not coupled to a heat
bath but obeyed pure Newtonian dynamics, while the walls were
thermostated. Note that in computing the local temperature at $z$
the streaming velocity $\vec{u}(z)$ is subtracted from the
instantaneous velocities of all particles in the interval [$z, z+
dz]$. From Varnik and Binder \cite{60}.}
\end{figure}

\begin{figure}
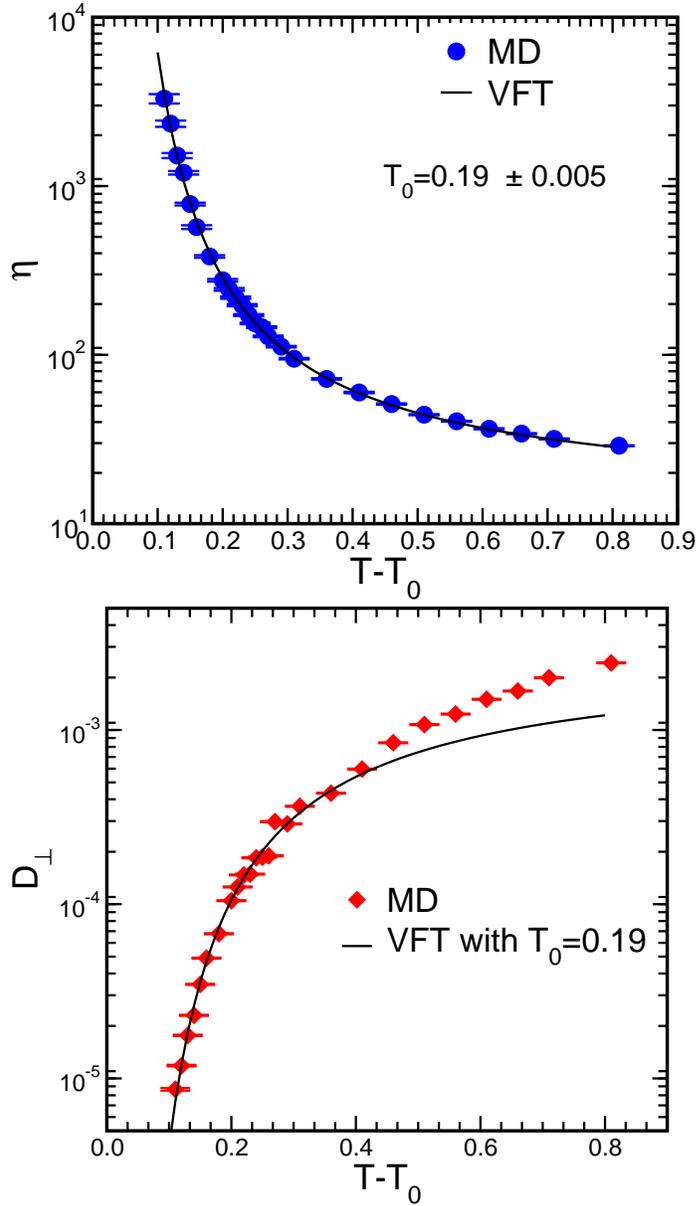
 
\hspace*{0.8cm}
\psfig{file=./fig10a.eps,height=8cm}

\hspace*{0.8cm}
\psfig{file=./fig10b.eps,height=8cm}
\vspace*{-0.3cm}
\caption{\label{fig10}
a) Shear viscosity of the model polymer melt as a function of the
temperature distance $T-T_0$ from the Vogel-Fulcher-Tammann
(VFT)-temperature $T_0=0.190 \pm 0.005$. Symbols denote the data
obtained for constant density $\rho_0=0.99$. The curve is a fit to
the VFT equation \{Eq.~(\ref{eq62})\}. b) Same as a) but for the
selfdiffusion constant $D_\bot$ (measured in the $y$-direction
transverse to the flow). From Varnik and Binder \cite{60}.}
\end{figure}

In this brief review, we have attempted to convey to the reader
the flavor of MD simulations, addressing both the estimation of
equilibrium properties such as static structure factors
(Fig.~\ref{fig4}) or density profiles near confining walls
(Fig.~\ref{fig8}a) and of dynamic properties, such as single chain
intermediate dynamic structure factors in polymer melts
(Figs.~\ref{fig1}, \ref{fig2}), diffusion constants in SiO$_2$
(Fig.~\ref{fig5}) and glass forming polymer melts
(Fig.~\ref{fig10}b), and collective transport coefficients such as
viscosity (Fig.~\ref{fig3},~\ref{fig10}a), thermal conductivity,
etc. Both type of MD simulations, dealing with systems in thermal equilibrium and
NEMD, has been discussed (Sec. 4).

\begin{figure} 
\psfig{file=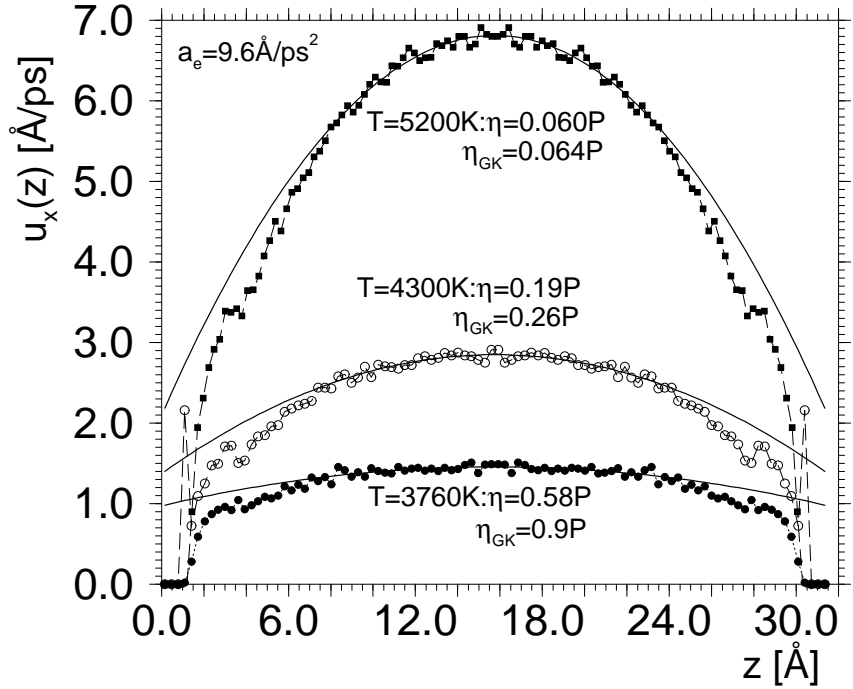,height=9cm}
\caption{\label{fig11}
Velocity profile of molten SiO$_2$ between atomistic walls, for an
acceleration $a_x=9.6$~\AA$/$ps$^2$, and three temperatures as
indicated. Points are NEMD data using a simulation box $L_x \times
L_y \times D$, with $L_x=L_y=23.0 $~\AA, $D=31.5$~ \AA, using
altogether $N=1152$ atoms. Curves are fits to Eq.~(\ref{eq60}).
The resulting viscosities are quoted in the figure, together with
the corresponding Green-Kubo estimates $\eta_{\rm GK}$ from
Ref. \cite{48}. From Horbach and Binder \cite{61}.}
\end{figure}

Already with respect to static properties, one has the choice
between different ensembles (microcanonical $NVE$ ensemble, or
$NVT$ and $NpT$ ensembles, realized with suitable thermostats and
barostats). It depends on the questions which one likes to answer
which ensemble is more appropriate. Similarly, we have
demonstrated that transport coefficients (such as viscosity,
thermal conductivity, etc.) can be estimated from thermal
equilibrium simulations (via the Green-Kubo \cite{66} relations of
Sec.~2.4) or via NEMD work. There are many variants how one can
proceed \cite{20,21,56,57,58,59,60,61,62,65}, and clearly there is
no full consensus in the literature about the ``pros'' and
``cons'' of the various approaches yet. Instead the subject is
still a matter of current research. As an example,
Fig.~\ref{fig11} presents an example, again SiO$_2$, where
viscosities were estimated from a NEMD approach in complete
analogy with Fig.~\ref{fig8}b, and the data are compared to the
Green-Kubo approach used in Fig.~\ref{fig3}. At least within the
error bars noted in Fig.~\ref{fig3}, there is very good mutual
agreement, and this is reassuring because it shows that
the various systematic errors of the simulations and their
analysis are under control. We have emphasized that one must be
aware of systematic errors due to the finite size of the
simulation box, due to incomplete equilibration, in particular in
the case of long wavelength fluctuations, due to inaccuracies of
the algorithm related to the size of the time step, etc.

Finally, we emphasize that we have used illustrative examples
which were taken from the research of the authors for the sake of
simplicity only --- similarly valuable research on related problems
is available in the literature from many other groups, as is well
documented (see e.g.~Refs.~\cite{1,2,3,7,19,20,21}). We hope that
the present article ``wets the appetite'' of the reader to study
this extensive literature. \\

\bigskip
{\underline{\bf{Acknowledgments:}}} The research reviewed here
was supported by the Bundesministerium f\"ur Bildung und Forschung
(BMBF No 03N6500 and 03N6015), by the Deutsche
Forschungsgemeinschaft (DFG), grants SFB/262/D1, D2, HO 2231/2-1 and
by SCHOTT GLAS. We are very grateful to J. Baschnagel, C.
Bennemann, A. Latz, P. Scheidler, G. D. Smith, and K. Vollmayr-Lee for
their valuable collaboration on the research projects from which
these examples were taken. Generous grants of computing time 
by the computer center of the University of Mainz (ZDV) and on
CRAY-T3E from NIC J\"ulich and HLRS Stuttgart also is
acknowledged.

\clearpage

\clearpage

\end{document}